\documentclass[sigplan,nonacm,10pt]{acmart}
\hypersetup{draft}
\renewcommand{\comment}{3}{}

\usepackage{mathpartir}
\usepackage{thmtools}
\usepackage{thm-restate}
\usepackage{paralist}
\usepackage{xspace}
\usepackage{subcaption} %
\usepackage{multirow}
\usepackage{algorithm}
\usepackage[noend]{algpseudocode}
\usepackage{dirtree}
\usepackage{listings}
\usepackage{bm}
\usepackage{siunitx}
\usepackage{extarrows}
\usepackage{stmaryrd}
\usepackage{tabularx}
\usepackage{microtype}
\usepackage{balance}

\usepackage{tikz}
\usetikzlibrary{shapes.arrows}
\usetikzlibrary{positioning,automata,arrows}

\pgfdeclareimage[width=6em]{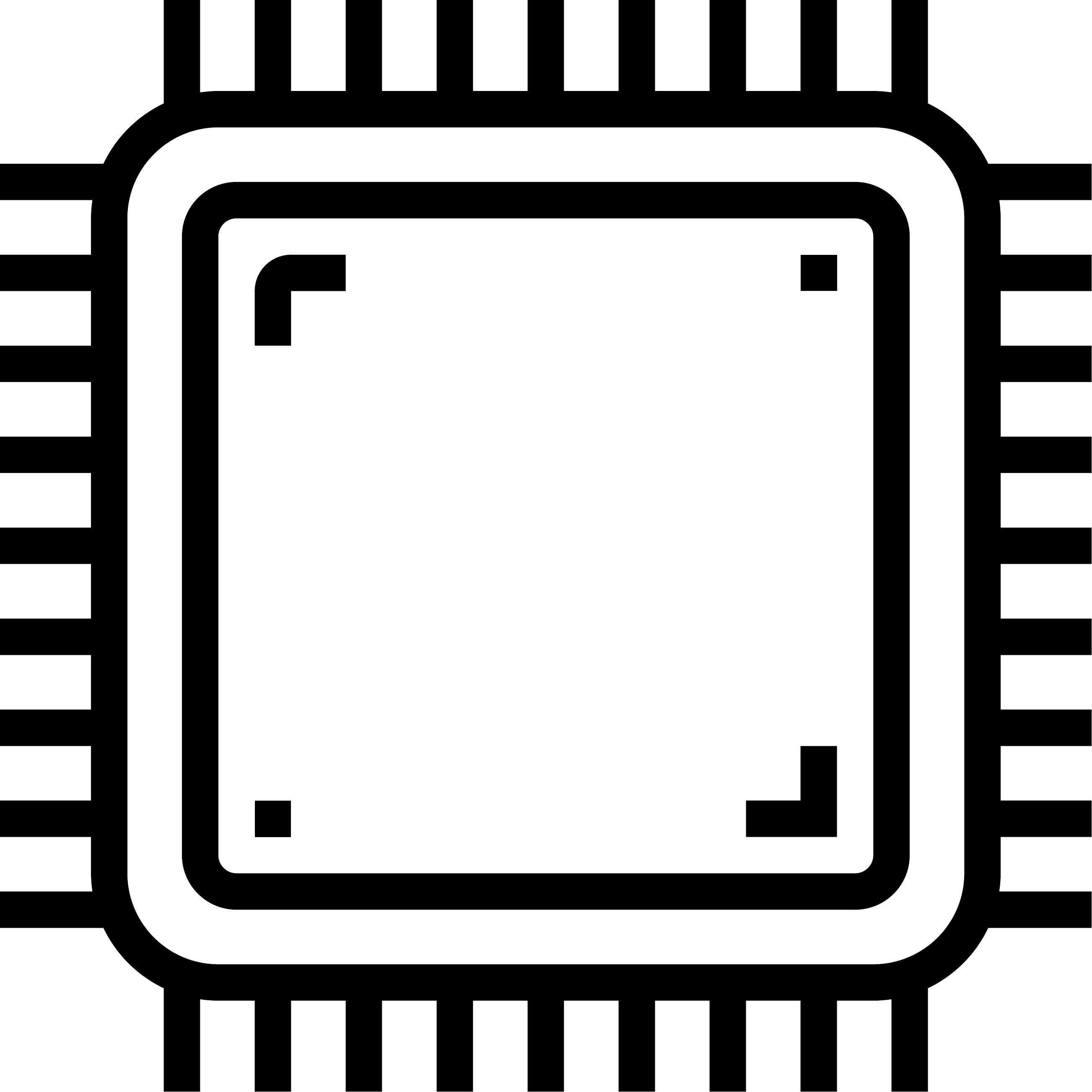}{processor}

\definecolor{Blue3}{HTML}{0000CD}
\definecolor{Green4}{HTML}{008B00}
\definecolor{Red3}{HTML}{CD0000}
\definecolor{orange}{rgb}{0.8, 0.47, 0.196}
\lstdefinestyle{Cstyle}
{
	frame = tb,
  	showstringspaces = false,
  	breaklines = true,
  	breakatwhitespace = true,
  	tabsize = 3,
    language = {[ANSI]C},
    alsoletter={.\$},
    basicstyle={\small\ttfamily\color{black}},
    stringstyle={\ttfamily\color{string-color}},
    keywordstyle={\ttfamily\color{Blue3}},
    keywordstyle=[2]{\ttfamily\color{Green4}},
    keywordstyle=[3]{\ttfamily\color{orange}},
    keywordstyle=[4]{\ttfamily\color{violet}},
    otherkeywords = {cache, age, content, HIT, MISS, access, hit, miss, bit, Block, policy, not},
    morekeywords = [2]{age,content, P, C, P},
    morekeywords = [3]{access,hit,miss,replacement,promote, normalize, evict, insert},
    morekeywords = [4]{HIT, MISS},
}
\newcommand{\inlineCode}[1]{\lstinline[style=Cstyle]|#1|}

\newcolumntype{L}[1]{>{\raggedright\let\newline\\\arraybackslash\hspace{0pt}}m{#1}}

\definecolor{ao(english)}{rgb}{0.0, 0.5, 0.0}
\definecolor{royalblue(web)}{rgb}{0.25, 0.41, 0.88}

\renewcommand{\comment}[2][1=]{%
	\colorlet{colorVar}{red}%
	\setCommentColor{#1}%
	\textcolor{colorVar}{[\commentAuthor{#1}#2]}}

\newcommand{\setCommentColor}[1]{%
	\ifthenelse{\equal{#1}{bk}}%
		{\colorlet{colorVar}{red!50}}%
		{\ifthenelse{\equal{#1}{pv}}%
			{\colorlet{colorVar}{blue}}%
			{\ifthenelse{\equal{#1}{mg}}%
				{\colorlet{colorVar}{ao(english)}}%
			{\ifthenelse{\equal{#1}{pg}}%
				{\colorlet{colorVar}{magenta}}%
				{}%
			}%
		}%
	}%
}

\newcommand{\commentAuthor}[1]{%
	\ifthenelse{\equal{#1}{bk}}%
		{Boris:\ }%
		{\ifthenelse{\equal{#1}{pv}}%
			{Pepe:\ }%
			{\ifthenelse{\equal{#1}{mg}}%
				{Marco:\ }%
			{\ifthenelse{\equal{#1}{pg}}%
				{Pierre:\ }%
				{}%
			}%
		}%
	}%
}

\mathchardef\breakingcomma\mathcode`\,
{\catcode`,=\active
  \gdef,{\breakingcomma\discretionary{}{}{}}
}
\newcommand{\activeComma}[1]{\mathcode`\,=\string"8000 #1}

\newcommand{\col}[2]{\ensuremath{{\color{#1}{#2}}}}

\newcommand{\mtt}[1]{\ensuremath{\mathtt{#1}}}
\newcommand{\mf}[1]{\ensuremath{\mathbf{#1}}}

\newcommand{\cache}[1]{\mtt{\col{ao(english)}{#1}}}
\newcommand{\policy}[1]{{\mf{\col{royalblue(web)}{#1}}}}

\newcommand{\Nat}{\mathbb{N}}

\newcommand{\Blocks}{\mathbb{B}}
\newcommand{\CacheBlocks}{\cache{Blocks}}
\newcommand{\CacheHit}{\cache{Hit}}
\newcommand{\CacheMiss}{\cache{Miss}}
\newcommand{\PolicyHit}[1]{\policy{Ln(#1)}}
\newcommand{\PolicyMiss}{\policy{Evct}}
\newcommand{\PolicyEvict}[1]{\policy{#1}}%
\newcommand{\PolicyNoEvict}{\policy{\bot}}%

\newcommand{\sizeof}[1]{\left| #1 \right|}

\algdef{SE}[DOWHILE]{Do}{doWhile}{\algorithmicdo}[1]{\algorithmicwhile\ #1}%

\newcommand{\cachequery}{\textsc{CacheQuery}}
\newcommand{\tool}{\textsc{Polca}}
\newcommand{\mapper}{\textsc{Polca}}

\newcommand{\tup}[1]{\langle \activeComma{#1} \rangle}
\newcommand{\ControlStates}{\policy{CS}}
\newcommand{\controlState}{\policy{cs}}
\newcommand{\policySemantics}[1]{ \policy{\llbracket} {#1} \policy{\rrbracket}}
\newcommand{\cacheSemantics}[1]{ \cache{\llbracket} {#1} \cache{\rrbracket}}
\newcommand{\semantics}[1]{\llbracket #1 \rrbracket}
\newcommand{\CacheContents}[1]{\cache{{CC}}^{#1}}
\newcommand{\cacheContent}{\cache{cc}}
\newcommand{\oracleCache}{\mathit{probeCache}}

\newcommand{\evalCache}[1]{\cache{\xLongrightarrow{#1}}}
\newcommand{\evalPol}[1]{\policy{\xrightarrow{#1}}}
\newcommand{\element}[2]{#1[#2]}

\newcommand{\update}[3]{#1[#2 \mapsto #3]}

\newcommand{\finiteSequences}[1]{#1^*}
\newcommand{\emptysequence}{\varepsilon}
\newcommand{\concat}{\cdot}

\newcommand{\io}[2]{\tup{#1,#2}}
\newcommand{\PolicyIo}[2]{\policy{\tup{#1,#2}}}
\newcommand{\CacheIo}[2]{\cache{\tup{#1,#2}}}
\newcommand{\para}[1]{\paragraph{#1.}}
\newcommand{\inputPolicy}{\policy{IP}}
\newcommand{\outputPolicy}{\policy{OP}}
\newcommand{\inputCache}{\cache{IC}}
\newcommand{\outputCache}{\cache{OC}}

\newcommand{\PolicyTransitionFunction}{\policy{\delta}}
\newcommand{\PolicyOutputFunction}{\policy{\lambda}}
\newcommand{\Cache}[3]{\cache{C(}#1,#2,#3\cache{)}}

\begin{document}

\title[CacheQuery: Learning Replacement Policies from Hardware Caches]{CacheQuery: Learning Replacement Policies\texorpdfstring{\\}{ }from Hardware Caches}

\author{Pepe Vila}
\affiliation{%
  \institution{IMDEA Software Institute}
  \institution{Universidad Polit{\'e}cnica de Madrid}
  \country{Spain}
}

\author{Pierre Ganty}
\affiliation{%
  \institution{IMDEA Software Institute}
  \country{Spain}
}

\author{Marco Guarnieri}
\affiliation{%
 \institution{IMDEA Software Institute}
\country{Spain}
}

\author{Boris K\"opf}
\affiliation{%
  \institution{Microsoft Research}
  \country{United Kingdom}
}

\begin{abstract}
	We show how to infer deterministic cache replacement policies using off-the-shelf automata learning and program synthesis techniques. For this, we construct and chain two abstractions that expose the cache replacement policy of any set in the cache hierarchy as a membership oracle to the learning algorithm, based on timing measurements on a silicon CPU. Our experiments demonstrate an advantage in scope and scalability over prior art and uncover two previously undocumented cache replacement policies.

\end{abstract}




\maketitle
\pagestyle{plain}

%
% start input ./introduction.tex
%
\section{Introduction}

Understanding the timing behavior of modern CPUs is crucial for optimizing code and for ensuring timing-related security and safety properties. Examples of such properties are bounds on programs' worst-case execution time~\cite{wcetsurvey08} or on the amount of information leaked via timing~\cite{cacheaudit-tissec,brotzman2019casym}.
Unfortunately, the timing behavior of today's high-performance processors depends on subtle and poorly documented details of their microarchitecture, which has triggered laborious efforts to build models of different components~\cite{nanobench2019,Irazoqui15systematicreverse,MauriceSNHF15,drama2016}.

Cache replacement policies have received specific attention~\cite{grueda13,cispa596,ReloadRefresh2019,ReplacementPolicy2013,AbelISPASS214}, because they control the content of the memory hierarchy and hence heavily influence execution time. Detailed policy models are used in worst-case execution time analyzers~\cite{absint}, CPU simulators~\cite{binkert2011gem5}, and for improving microarchitectural attacks and defenses~\cite{ReloadRefresh2019,cacheaudit-tissec,brotzman2019casym}.

However, only few authors have approached the problem of inferring replacement policies in a principled way.\looseness=-1
\begin{asparaitem}
    \item Rueda~\cite{grueda13} uses off-the-shelf techniques for learning register automata to infer cache replacement policies. The approach learns replacement policies with small state-spaces from noiseless simulator traces, but it has not been successfully applied to actual hardware.
	\item Abel and Reineke~\cite{cispa596} present an approach that infers so-called permutation-based replacement policies, which include LRU, FIFO, and PLRU~\cite{Handy1993}. The approach has been used to infer policies from hardware performance counter measurements
	on hardware. However, permutation-based policies are restrictive in that they do not include important examples such as MRU~\cite{mru1992}, SRRIP~\cite{Jaleel2010}, or the policies implemented in the lower-level caches of recent Intel CPUs.
\end{asparaitem}

Furthermore, both approaches share a drawback: the inferred policies are not easily interpretable by humans.

%
% start input ./figure.tex
\begin{figure*}[ht]
\newcommand{\arrowLength}{25mm}
\newcommand{\arrowWidth}{.5mm}
\newcommand{\arrowExtend}{1mm}
\begin{subfigure}[t]{.25\textwidth}
\centering
\begin{tikzpicture}
\tikzstyle{every state}=[fill=none,draw=black,text=black, inner sep=0pt,outer sep = 0pt]
\tikzset{every state/.append style={rectangle}}

\draw (0,0) rectangle (1,2.5) node[pos=.5,rotate=90] (learnlib) {LearnLib~\cite{Steffen2011}};

\node[fill=green!50, single arrow,
		inner sep=\arrowWidth,minimum height=\arrowLength, single arrow head extend=\arrowExtend,
		rotate=0,  right=0.5cm of learnlib.south] (arrow1) {};
\node[fill=red!50, single arrow,
		inner sep=\arrowWidth,minimum height=\arrowLength, single arrow head extend=\arrowExtend,
		rotate=180, below=0.3cm of arrow1] (rarrow1) {};
\node[above=0.1cm of arrow1] {
	\begin{tabular}{c}
	\footnotesize{}
	\footnotesize{$\PolicyHit{0}~\PolicyHit{1}~\PolicyMiss$} \\
	\end{tabular}
};
\node[below=0.2cm of rarrow1] {
	\begin{tabular}{c}
	\footnotesize{$\PolicyNoEvict~\PolicyNoEvict~\PolicyEvict{0}$} \\
	\footnotesize{}
	\end{tabular}
};

\end{tikzpicture}
\caption{LearnLib~\cite{Steffen2011} issues membership queries to the system under learning (SUL). The salient feature of our approach is that the language of the SUL refers to cache lines and {\em not} to the cache content. When the learning loop terminates, our tool returns an automaton describing the cache replacement policy under learning.} 
\label{figure:overview:learnlib}
\end{subfigure}
\hspace{1em}
\begin{subfigure}[t]{.30\textwidth}
\centering
\begin{tikzpicture}
\tikzstyle{every state}=[fill=none,draw=black,text=black, inner sep=0pt,outer sep = 0pt]
\tikzset{every state/.append style={rectangle}}

\draw[] (0,0) rectangle (1,2.5) node[pos=.5,rotate=90] (interpol) {\mapper{}};

\node[fill=green!50, single arrow,
		inner sep=\arrowWidth,minimum height=\arrowLength, single arrow head extend=\arrowExtend,
		rotate=0,  right=1cm of learnlib.south] (arrow1) {};
\node[fill=red!50, single arrow,
		inner sep=\arrowWidth,minimum height=\arrowLength, single arrow head extend=\arrowExtend,
		rotate=180, below=0.3cm of arrow1] (rarrow1) {};
\node[above=0.1cm of arrow1] {
	\begin{tabular}{c}
	\footnotesize{$\cache{A}~\cache{B}~\cache{C}~\cache{A}$} \\
	\footnotesize{$\cache{A}~\cache{B}~\cache{C}~\cache{B}$}
	\end{tabular}
};
\node[below=0.15cm of rarrow1] {
	\begin{tabular}{c}
	\footnotesize{$\CacheHit~\CacheHit~\CacheMiss~\CacheMiss$} \\
 	\footnotesize{$\CacheHit~\CacheHit~\CacheMiss~\CacheHit$}
	\end{tabular}
};
\end{tikzpicture}
\
\caption{\mapper{} translates a sequence of requests for cache lines $\PolicyHit{i}$ or evictions $\PolicyMiss$ into sequences of abstract memory blocks. For this, the algorithm keeps track of the current cache state (here: blocks \cache{A}/\cache{B} in lines \policy{0}/\policy{1}). $\PolicyMiss$ spawns multiple sequences that first produce a cache miss (here: \cache{C}), followed by accesses to all previously contained blocks to infer which line was evicted (here: \policy{0}).}
\label{figure:overview:interpol}
\end{subfigure}
\hspace{1em}
\begin{subfigure}[t]{.35\linewidth}
\centering
\begin{tikzpicture}
\tikzstyle{every state}=[fill=none,draw=black,text=black, inner sep=0pt,outer sep = 0pt]
\tikzset{every state/.append style={rectangle}}

\draw[] (0,0) rectangle (1,2.5) node[pos=.5,rotate=90] (cachequery) {\cachequery{}};

\node[fill=green!50, single arrow,
		inner sep=\arrowWidth,minimum height=\arrowLength, single arrow head extend=\arrowExtend,
		rotate=0,  right=0.5cm of learnlib.south] (arrow1) {};
\node[fill=red!50, single arrow,
		inner sep=\arrowWidth,minimum height=\arrowLength, single arrow head extend=\arrowExtend,
		rotate=180, below=0.3cm of arrow1] (rarrow1) {};
\node[above=0.1cm of arrow1] {
	\begin{tabular}{c}
	\footnotesize{\texttt{f30 f40 f50 f30}} \\
	\footnotesize{\texttt{f30 f40 f50 f40}} \\
	\end{tabular}
};
\node[below=0.2cm of rarrow1] {
	\begin{tabular}{c}
	\footnotesize{4c 4c 12c 12c} \\
	\footnotesize{4c 4c 12c 4c} \\
	\end{tabular}
};

\node[right=0.1cm of arrow1] (processor) {\pgfuseimage{processor}};

\end{tikzpicture}
\caption{\cachequery{} receives as input sequences of abstract blocks (e.g., \cache{A~B~C~A}) and translates them into distinct concrete memory blocks (e.g., \cache{A} maps to address \texttt{f30}) that all map into the same cache set. It loads the corresponding memory blocks, counts the corresponding clock cycles, and returns for each load whether it was a cache hit (e.g., the \texttt{4c} measurement maps to \CacheHit) or a miss.
}
\label{figure:overview:cachequery}
\end{subfigure}
\caption{Leveraging \mapper{} and \cachequery{} to learn a toy replacement policy of a 2-way set associative CPU cache.}%
\label{figure:overview}
\end{figure*} % end input ./figure.tex
 
\para{Approach}
In this paper we propose an approach for learning cache replacement policies that goes beyond the state-of-the-art in that it (1) can learn arbitrary deterministic  policies (2) from real-time (or performance counter) measurements on silicon CPUs. Moreover, we show how to (3) apply program synthesis to yield human-readable interpretations of the inferred policies.

Our approach relies on two contributions that enable us to leverage off-the-shelf automata learning tools~\cite{learnlib,Steffen2011} for attacking the problem:

\begin{asparaitem}
	\item A tool, called \cachequery{}, that provides an abstract interface to any {\em individual cache set} within the cache hierarchy\footnote{For a primer on hardware caches, see~\S~\ref{sec:model:primer}.} of a silicon CPU. With \cachequery{}, users can specify a cache set (say: set 63 in the L2 cache) and a pattern of memory accesses (say: \texttt{A B C A B C}), and they receive as output a sequence (say: \texttt{Miss Miss Miss Hit Hit Hit}) representing the hits and misses produced when performing a sequence of memory loads to addresses that are mapped into the specified cache set and that follow the specified pattern. \cachequery{} liberates the user from dealing with intricate details such as the virtual-to-physical memory mapping, cache slicing, set indexing, interferences from other levels of the cache hierarchy, and measurement noise, and thus enables civilized interactions with an individual cache set. See Figure~\ref{figure:overview:cachequery}.\looseness=-1

	\item An algorithm, called \mapper{}, that provides an abstract interface to the {\em cache replacement policy} based on an interface to a cache set, such as \cachequery{}. \mapper{} translates inputs to the replacement policy (which refer to the cache {\em lines}) into inputs to the cache set (which refer to the {\em memory blocks} that are stored in it). To achieve this, \mapper{} itself keeps track of the current cache content, e.g., by issuing queries to the cache interface to determine which block has been evicted in a miss.  \mapper{} exploits the data-independence symmetry of the replacement policy that would otherwise have to be inferred by the learning algorithm, and it is key to making automata learning work in this domain. See Figure~\ref{figure:overview:interpol}.
\end{asparaitem}

We use \mapper{} as a so-called {\em membership oracle} for the replacement policy to be learned from an abstract cache set (which can be implemented by \cachequery{} or by a software-simulated cache). The oracle provides an interface to libraries such LearnLib~\cite{Steffen2011}, which enables us to leverage the state-of-the-art in automata learning for inferring replacement policies of silicon CPUs. We give formal relative completeness guarantees for the learned policies, based on the correctness of \mapper{} and LearnLib. The full learning pipeline is depicted in Figure~\ref{figure:overview}.

Finally, we show how to use program synthesis to automatically derive a higher-level representation of the learned replacement policy.
The main component of our synthesis step is a \textit{program template} for replacement policies, which we base on concepts used to describe replacement policies in the microarchitecture community~\cite{Jaleel2010}.
By combining the policy template with a set of constraints derived from the learned automaton, we can rely on off-the-shelf synthesis solvers to synthesize high-level policy representations~\cite{sketch2009}.

\para{Evaluation}
We evaluate our approach in 3 case studies:
\begin{asparaenum}
\item We evaluate the scalability of learning replacement policies with \mapper{}. To this end, we learn a comprehensive set of deterministic  policies (including FIFO, LRU, PLRU~\cite{Handy1993}, MRU~\cite{mru1992}, LIP~\cite{Qureshi2007}, and different variants of SRRIP~\cite{Jaleel2010}) from the noiseless hit-miss traces produced by a software-simulated cache. Our experiments demonstrate that \mapper{} enables LearnLib to infer policies with state-spaces of more than 2 orders of magnitude larger than what was reported for direct applications of LearnLib to simulator traces~\cite{grueda13}.
\item We evaluate the effectiveness of learning with \mapper{} and \cachequery{} on the L1, L2, and L3 caches of three recent Intel CPUs. %
Our experiments show that we can effectively learn cache replacement policies up to associativity 8 from timing measurements on modern CPUs. While some of the policies were known, we do uncover 2 policies that have not yet been documented in the literature.
\item We evaluate our template-based synthesis approach by synthesizing programs for 8 out of 9 different policies (obtained from both the simulators and silicon CPUs), for a fixed associativity 4. This allows us to provide high-level descriptions for the 2 previously undocumented policies.
\end{asparaenum}

\para{Summary of Contributions}
In summary, we present a practical end-to-end solution for inferring deterministic cache replacement policies using off-the-shelf techniques for automata learning and program synthesis. The enabling contribution is a chain of two abstractions that exposes a membership oracle to the cache replacement policy, based on timing measurements on a silicon CPU.

Our tools, \cachequery{} and \tool{}, are available, together with the learned models and synthesized programs, at \url{https://github.com/cgvwzq/cachequery/} and \url{https://github.com/cgvwzq/polca/} respectively.
Persistently archived full repositories for our tools are available at~\cite{swhpolca,swhcachequery}.
%

%
%
%
%
%
%
%
%

 % end input ./introduction.tex
 %
%
% start input ./model.tex
%
\section{Modeling Caches and Policies}\label{sec:model}

In this section we present our model of hardware caches.
A key feature, inspired by~\cite{PabloIncomparability2019}, is that we distinguish between the replacement policy (\S~\ref{sec:formal-model:replacement-policies}), which determines the cache lines to be replaced, and the cache itself (\S~\ref{sec:formal-model:caches}), which stores and retrieves memory blocks (according to the replacement policy).
We start by introducing necessary background.

\subsection{A Primer on Hardware Caches}\label{sec:model:primer}

Caches are fast but small memories that bridge the latency gap between the CPU and the main memory. To profit from spatial locality and to reduce management overhead, the main memory is logically partitioned into a set of {\em blocks}.

Each block is cached as a whole in a cache {\em line} of the same size. When accessing a block, the cache logic determines whether the block is stored in the cache (a cache hit) or not (a cache miss).
For this purpose, caches are partitioned into equally sized cache {\em sets}. The capacity of a set-associative cache set is called {\em associativity} (or {\em ways}) and represents the number of lines per set.
Because the cache is smaller than the main memory, upon a cache miss, a {\em replacement policy} must decide which memory block to {\em evict} in order to make room for the more recent block.

In this work, we consider $n$-way set associative caches, i.e., caches where all cache sets consist of $n$ lines, and we focus on individual cache sets.
For brevity's sake, in the following we refer to a  set of an $n$-way cache simply as an $n$-way cache.

\subsection{Replacement Policy Model}\label{sec:formal-model:replacement-policies}

We model the replacement policy of a cache set as a deterministic, finite-state Mealy machine that accepts inputs of the form  $\PolicyHit{i}$, for accessing the $i$-th cache line, and  $\PolicyMiss$, for requesting a cache line to be freed (cf.~Table~\ref{tab:alphabets}).
Given an input, the policy updates its control state and outputs the index of the line to be freed (or \(\PolicyNoEvict\) otherwise).

\begin{table}[h]
	\centering
	\caption{Policy and cache alphabets (associativity $n$)}
	\label{tab:alphabets}
\begin{tabular}{l | c c}
& \textbf{Policy} & \textbf{Cache} \\
\hline
Input & \( \{ \PolicyHit{0},\ldots,\PolicyHit{n{-}1}\}\cup \{ \PolicyMiss \}\) & \(\CacheBlocks\) \\
	Output &
\( \{ \PolicyNoEvict \}\cup \{ \PolicyEvict{0},\ldots,\PolicyEvict{n-1} \}\)
	& \(\{\CacheHit, \CacheMiss\}\)
\end{tabular}
\end{table}

\begin{definition}\label{def:policy}
	A {\em replacement policy} of associativity $n \in \Nat$ is a Mealy machine $\tup{\ControlStates, \policy{\controlState_0}, \inputPolicy, \outputPolicy, \PolicyTransitionFunction, \PolicyOutputFunction}$ consisting of:
\begin{asparaitem}
\item a finite set of {\em control states} $\ControlStates$;
\item an {\em initial} control state $\policy{\controlState_0} \in \ControlStates$;
\item the set of inputs $\inputPolicy = \{ \PolicyHit{0}, \ldots, \PolicyHit{n-1}\} \cup \{ \PolicyMiss \}$; %
\item the set of outputs $\outputPolicy = \{ \PolicyNoEvict\} \cup \{ \PolicyEvict{0},\ldots,\PolicyEvict{n-1} \}$;%
\item a transition function $\PolicyTransitionFunction : \ControlStates \times \inputPolicy \to \ControlStates$; and %
\item an output function $\PolicyOutputFunction:  \ControlStates \times \inputPolicy \to \outputPolicy$. %
\end{asparaitem}
We require that, (a) \(\PolicyOutputFunction\) returns a value in \(\{\PolicyEvict{0},\ldots,\PolicyEvict{n-1}\}\) when given the input \(\PolicyMiss\); and (b)
\(\PolicyOutputFunction\) returns
\(\PolicyNoEvict\) when given an input in \(\{ \PolicyHit{0},\ldots,\PolicyHit{n{-}1} \}\).
\end{definition}
We write $\controlState \evalPol{\io{\policy{i}}{\policy{o}}} \policy{\controlState'}$ when $\PolicyTransitionFunction(\controlState,\policy{i}) = \policy{\controlState'}$ and $ \PolicyOutputFunction(\controlState,\policy{i}) = \policy{o}$.

We now introduce the trace semantics of policies, where traces are sequences of input/output pairs.
We use standard sequence notation:
$\finiteSequences{S}$ is the set of finite sequences over $S$, $\emptysequence$ is the  empty sequence, and $s_1 \concat s_2$ denotes sequence concatenation.
The \textit{trace semantics of a policy} $\policy{P}$ (short: {\em policy semantics}) %
is the set $\policySemantics{\policy{P}} \subseteq \finiteSequences{(\inputPolicy \times \outputPolicy)}$ of all sequences $\PolicyIo{i_1}{o_1} \concat \PolicyIo{i_2}{o_2} \concat \ldots \concat \PolicyIo{i_m}{o_m}$ for which
there are control states $\policy{\controlState_1}, \ldots, \policy{\controlState_{m}}$ such that
$\policy{\controlState_0} \evalPol{\PolicyIo{i_1}{o_1}} \policy{\controlState_1} \evalPol{\PolicyIo{i_2}{o_2}} \ldots \evalPol{ \PolicyIo{i_m}{o_m}} \policy{\controlState_{m}}$.

\begin{example}
\label{example:lru-policy}
Consider a \textit{Least Recently Used} (LRU) replacement policy, where the least recently used cache line is the one to be evicted.
The LRU policy with associativity 2 can be formalized with the following Mealy Machine: %
{
\begin{center}
\begin{tikzpicture}[->,>=stealth',shorten >=1pt,auto,node distance=2.8cm,
                    semithick]
  \tikzstyle{every state}=[fill=none,draw=black,text=black]
  \node[state] (A)                    {$\policy{\controlState_0}$};
  \node[state] (B) [right of=A]       {$\policy{\controlState_1}$};

  \path (A) edge [loop left] node {$\PolicyIo{\PolicyHit{1}}{\PolicyNoEvict}$} (A)
            edge [bend right, anchor=center, below] node {$\PolicyIo{\PolicyHit{0}}{\PolicyNoEvict},\PolicyIo{\PolicyMiss}{\PolicyEvict{0}}$} (B)
        (B) edge [loop right] node {$\PolicyIo{\PolicyHit{0}}{\PolicyNoEvict}$} (B)
            edge [bend right,  anchor=center, above]  node {$\PolicyIo{\PolicyHit{1}}{\PolicyNoEvict},\PolicyIo{\PolicyMiss}{\PolicyEvict{1}}$} (A)
;
\end{tikzpicture}
\end{center}
}

\noindent
There are two control states $\policy{\controlState_0}$ and $\policy{\controlState_1}$, where \(\policy{\controlState_i}\) indicates that line $\policy{i}$ is next to be evicted (i.e., $\policy{i}$ is the line storing the least recently used memory block).
\end{example}

\subsection{Cache Model}\label{sec:formal-model:caches}

We model a cache of associativity $n$ as a Labeled Transition System (LTS) that accepts as input elements $\cache{b}$ from a potentially infinite set of memory blocks $\CacheBlocks$, and that produces as output a \(\CacheHit\) when block $\cache{b}$ is in the cache, and a \(\CacheMiss\) otherwise (cf.~Table~\ref{tab:alphabets}).

Each state of the cache is a pair $\tup{\cacheContent,\controlState}$ consisting of the cache content $\cacheContent\in \CacheContents{n}$, which is an $n$-tuple of memory blocks without repetitions, and the control state $\controlState$ of a replacement policy \policy{P}. Formally:

\begin{definition}\label{def:cache}
	An {\em $n$-way cache} induced by a policy $\policy{P} = \tup{\ControlStates, \policy{\controlState_0}, \inputPolicy, \outputPolicy, \PolicyTransitionFunction,\PolicyOutputFunction}$ is an LTS $\Cache{\policy{P}}{\cache{\cacheContent_0}}{n} = \tup{\cache{S}, \cache{s_0}, \inputCache, \outputCache, \mathord{\evalCache{}}}$ consisting of:

\begin{asparaitem}
\item a set of {\em cache states} $\cache{S} = \CacheContents{n} \times \ControlStates$; %
\item an {\em initial} cache state $\cache{s_0} = \tup{ \cache{\cacheContent_0}, \policy{\controlState_0}} \in \cache{S}$;
\item a set of inputs $\inputCache = \CacheBlocks$;
\item a set of outputs $\outputCache = \{\CacheHit, \CacheMiss \}$; %
\item a transition relation $\mathord{\evalCache{}} \subseteq \cache{S} \times \inputCache \times \outputCache \times \cache{S}$  that is induced by the policy $\policy{P}$ following Figure~\ref{fig:cache-rules}.
\end{asparaitem}
\end{definition}

\begin{figure}[h]%
	{
	\begin{mathpar}
	\inferrule* [right=Hit]
	{
		\element{\cacheContent}{\policy{i}} = \cache{b}\\
		\controlState \evalPol{\io{\PolicyHit{i}}{\PolicyNoEvict}} \policy{\controlState'}
	}
	{
	\tup{\cacheContent,\controlState} \evalCache{\io{\cache{b}}{\CacheHit}} \tup{\cacheContent, \policy{\controlState'}}
	}

	\inferrule* [right=Miss]
	{
		\forall \policy{i}.\ \element{\cacheContent}{\policy{i}} \neq \cache{b}\\
		\controlState \evalPol{\io{\PolicyMiss}{\PolicyEvict{i}}} \policy{\controlState'}
	}
	{
	\tup{\cacheContent,\controlState} \evalCache{\io{\cache{b}}{\CacheMiss}} \tup{\update{\cacheContent}{\policy{i}}{\cache{b}}, \policy{\controlState'}}
	}
	\end{mathpar}
	}
	\caption{Transition relation for a cache $\evalCache{}$ given that of a replacement policy $\evalPol{}$. Here, $\element{\cacheContent}{\policy{i}}$ denotes the block stored in $\cacheContent$'s $\policy{i}$-th line and $\update{\cacheContent}{\policy{i}}{\cache{b}}$ the cache content obtained by replacing the block in the $\policy{i}$-th line with $\cache{b}$.}
\label{fig:cache-rules}
\end{figure}

The cache's transition relation relies on two rules, see Fig.~\ref{fig:cache-rules}:
\looseness=-1

The rule \textsc{Hit} captures what happens upon access to a block that is cached:
The rule
(1) determines that  $\cache{b}$ is stored in $\cacheContent$'s $\policy{i}$-th line, and
(2) updates the control state by executing the policy with input $\PolicyHit{i}$. %

The rule \textsc{Miss} captures what happens upon access to a block that is not cached:
The rule
(1) checks that the block $\cache{b}$ is not in the cache,
(2) determines the line $\policy{i}$ of the block to evict by executing the policy with input $\PolicyMiss$, and %
(3) inserts $\cache{b}$ in the $\policy{i}$-th line and updates the cache state.

Note that the cache's transition relation directly updates only the cache content $\cacheContent$.
Changes to the control state $\controlState$ are mediated by the replacement policy, which takes as input only accesses to cache lines $\PolicyHit{i}$ or eviction requests $\PolicyMiss$.
Hence, updates to the control state are agnostic to the accessed memory blocks.

Similarly to a policy's  semantics, we introduce a cache's  semantics. The \textit{trace semantics of a cache} $\cache{C}$ (short: {\em cache semantics}) %
 is the set  $\cacheSemantics{\cache{C}} \subseteq \finiteSequences{(\inputCache \times \outputCache)}$ of all sequences $\CacheIo{i_1}{o_1} \concat \CacheIo{i_2}{o_2} \concat \ldots \concat \CacheIo{i_m}{o_m}$ for which
 there are cache states $\cache{s_1}, \ldots, \cache{s_{m}}$ such that
 $\cache{s_0} \evalCache{\CacheIo{i_1}{o_1}} \cache{s_1} \evalCache{\CacheIo{i_2}{o_2}} \ldots \evalCache{ \CacheIo{i_m}{o_m}} \cache{s_{m}}$.

\begin{example}\label{example:lru-cache}
	The LTS of the cache induced by the LRU policy from Example~\ref{example:lru-policy} is as follows (we depict only part of the infinite LTS for \(3\) abstract blocks \cache{A}
, \cache{B}, and \cache{C}):
\begin{center}
\begin{tikzpicture}[->,>=stealth',shorten >=1pt,auto,node distance=1.9cm, semithick]
  \tikzstyle{every state}=[fill=none,draw=black,text=black,rounded corners]
  \tikzset{every state/.append style={rectangle, rounded corners}}

  \node[state] (B)        {$\tup{\cache{\tup{\cache{A},\cache{B}}},\policy{\controlState_0}}$};
  \node[state] (A) [right=2cm of B]                   {$\tup{\cache{\tup{\cache{C},\cache{B}}},\policy{\controlState_1}}$};
  \node[state] (C) [below of=B]       {$\tup{\cache{\tup{\cache{A},\cache{B}}},\policy{\controlState_1}}$};
  \node[state] (D) [right=2cm of C]      {$\tup{\cache{\tup{\cache{A},\cache{C}}},\policy{\controlState_0}}$};

  \path (B) edge [loop left] node {$\CacheIo{\cache{B}}{\CacheHit}$} (B)
            edge [] node {$\CacheIo{\cache{C}}{\CacheMiss}$} (A)
            edge [bend right, anchor=center, left] node {$\CacheIo{\cache{A}}{\CacheHit}$} (C)
        (C) edge [loop left] node {$\CacheIo{\cache{A}}{\CacheHit}$} (C)
            edge [below]  node {$\CacheIo{\cache{C}}{\CacheMiss}$} (D)
            edge [bend right,  anchor=center, right]  node {$\CacheIo{\cache{B}}{\CacheHit}$} (B);
\end{tikzpicture}
\end{center}

\noindent
Consider the cache state $\tup{\cache{\tup{\cache{A},\cache{B}}}, \policy{\controlState_0}}$.
Accessing the block $\cache{B}$ produces a $\CacheHit$ since $\cache{B}$ is stored in line $1$.
Hence, we modify neither the cache content nor the control state because $\policy{\controlState_0} \evalPol{\io{\PolicyHit{1}}{\PolicyNoEvict}} \policy{\controlState_0}$ according to the policy.
Accessing the block $\cache{A}$, which is stored in line $0$, also produces a $\CacheHit$.
This time, however, we update the control state since the least recently used cache line is now $1$, i.e., $\policy{\controlState_0} \evalPol{\io{\PolicyHit{0}}{\PolicyNoEvict}} \policy{\controlState_1}$.
In contrast, accessing the block $\cache{C}$, which is not in the cache, leads to a $\CacheMiss$.
The replacement policy determines that the block $\cache{C}$ has to be stored in line $0$, i.e., $\policy{\controlState_0} \evalPol{\io{\PolicyMiss}{\PolicyEvict{0}}} \policy{\controlState_1}$, and the new cache state is $\tup{\cache{\tup{\cache{C},\cache{B}}},\policy{\controlState_1}}$.
\end{example}

%
%
%
%
%
%
%
%
%
%
%
%
%
%
%
%
%
%
%
%
%
%
%
%
%
%
%
%
%
%
%
%

%

 % end input ./model.tex
 %
%
% start input ./learning.tex
\section{\mapper{}: Learning Replacement Policies}\label{sec:learning}

In this section, we present our policy learning approach.
We begin by introducing background on automata learning (\S~\ref{sec:learning:overview}).
Next, we describe the two main components of our learning approach, namely oracles for membership (\S~\ref{sec:learning:membership}) and equivalence queries (\S~\ref{sec:learning:equivalence}) for replacement policies.
Finally, we describe our prototype implementation of \mapper{} on top of LearnLib (\S~\ref{sec:learning:implementation}).

\subsection{A Primer on Automata Learning}%
\label{sec:learning:overview}

The prevalent approach to learning automata follows the student-teacher paradigm established by Angluin~\cite{Angluin1987} and extended to Mealy machines by Niese~\cite{Niese2003}. There, the student's goal is to learn an unknown Mealy machine \(M\) by asking queries to the teacher.
There are two types of queries:
\begin{asparaenum}
	\item {\em membership queries}, where the student asks whether a given trace belongs to the machine \(M\), and
	\item {\em equivalence queries}, where the student asks whether a hypothesized Mealy machine \(H\) is (trace) equivalent to \(M\).
\end{asparaenum}

Initially, the student knows only the input and output alphabets. By making a finite number of queries to the teacher as prescribed by the learning algorithm, the student eventually learns \(M\).

We next show how to build oracles to answer membership and equivalence queries for a replacement policy, based on interactions with a hardware cache, which enables us to leverage automata learning algorithms for inferring replacement policies.

\subsection{Membership Queries}%
\label{sec:learning:membership}

We now present \mapper{}, an algorithm that provides a membership oracle for a replacement policy, given a cache that implements that policy. That is, \mapper{} takes as input a trace $\policy{t}$ of policy inputs and outputs,  a cache $\cache{C}$ induced by a policy $\policy{P}$, and it determines whether \(\policy{t} \in \policySemantics{\policy{P}}\).
For that, \mapper{} translates $\policy{t}$ into a series of input traces (i.e., sequences of memory blocks) to the underlying cache $\cache{C}$, by internally keeping track of the blocks stored in the cache. From the outputs of $\cache{C}$, \mapper{} then deduces whether $\policy{t} \in \policySemantics{\policy{P}}$.

By extracting the policy semantics $\policySemantics{\policy{P}}$ from the cache semantics $\cacheSemantics{\cache{C}}$, \mapper{} effectively inverts the transition rules in Figure~\ref{fig:cache-rules}.
This enables learning \textit{only} the policy $\policy{P}$, rather than learning also the data storage logic of the cache $\cache{C}$, and is key for scaling automata learning to hardware caches.\looseness=-1

\algrenewcommand\algorithmicindent{0.8em}%
\begin{algorithm}
\caption{\mapper{}: A membership oracle for policies}\label{alg:membership-query}
	\begin{algorithmic}[1]
		\Function{$\mapper{}$}{$\cache{\cacheContent_0}, \policy{t}, \cacheSemantics{\cache{C}}$}
		\State $\cacheContent \leftarrow \cache{\cacheContent_0}$ %
		\ForAll{$i=1,\ldots,|\policy{t}|$}
			\State let $\PolicyIo{ip}{op}$ be $\element{\policy{t}}{i}$
			\State \( \element{\cache{ic}}{i} \leftarrow \mathit{mapInput}(\policy{ip},\cacheContent) \) \label{lst:interpol:mapinput}
			\State \({\cache{oc}} \leftarrow \oracleCache(\cache{ic}[1 \ldots i], \cacheSemantics{\cache{C}})\) \label{lst:interpol:probecache}
			\State \( \policy{op'} \leftarrow \mathit{mapOutput}(\cache{oc}, \cache{ic}[1 \ldots i],\cacheContent,\cacheSemantics{\cache{C}}) \) \label{lst:interpol:mapoutput}
			\If{$\policy{op'} \neq \PolicyNoEvict$}
				\Comment{Update cache content}
				\State \(\cacheContent \leftarrow \update{\cacheContent}{\policy{op'}}{\element{\cache{ic}}{i}}\) \label{lst:interpol:update}
			\EndIf
			\If{$\policy{op} \neq \policy{op'}$} \label{lst:interpol:check}
				\State \Return $\mathit{false}$
			\EndIf
		\EndFor
		\State \Return $\mathit{true}$
	\EndFunction
	\Statex
	\Function{$\mathit{probeCache}$}{$\cache{q}, \cacheSemantics{\cache{C}}$}
		\State $k \leftarrow |\cache{q}|$
		\State let $\cache{o}$ be such that $\io{\element{\cache{q}}{1}}{\element{\cache{o}}{1}} \concat \ldots \concat  \io{\element{\cache{q}}{k}}{\element{\cache{o}}{k}} \in  \cacheSemantics{\cache{C}}$
		\State \Return $\element{\cache{o}}{k}$
	\EndFunction
	\Statex
	\Function{$\mathit{mapInput}$}{$\policy{ip},\cacheContent$}
		\If{$\policy{ip} \in \{\PolicyHit{0},\ldots,\PolicyHit{n-1}\}$}
			\State let $i$ be such that $\policy{ip} = \PolicyHit{i}$
			\State \Return $\element{\cacheContent}{i}$
			\Else  \Comment{$\policy{ip} = \PolicyMiss$}
			\State let $\cache{b} \in \CacheBlocks$ be such that \(\element{\cacheContent}{i} = \cache{b}\) for no \(i\)
			\State \Return $\cache{b}$
		\EndIf
	\EndFunction
	\Statex
	\Function{$\mathit{mapOutput}$}{$\cache{oc},\cache{q},\cacheContent,\cacheSemantics{\cache{C}}$}
		\If {\(\cache{oc} = \CacheHit \)}
			\State \Return \(\PolicyNoEvict\)
		\Else  \Comment{\(\cache{oc} = \CacheMiss \)}
			\State \Return \(\mathit{findEvicted}(\cache{q}, \cacheContent, \cacheSemantics{\cache{C}} )\) \label{lst:interpol:findevicted}
		\EndIf
	\EndFunction
	\Statex
	\Function{$\mathit{findEvicted}$}{$\cache{q}, \cacheContent, \cacheSemantics{\cache{C}}$}
		\ForAll{$i=1,\ldots,n$}
			\If{\(\oracleCache(\cache{q} \concat \element{\cacheContent}{i},\cacheSemantics{\cache{C}}) = \CacheMiss\)}
				\State \Return i
			\EndIf
		\EndFor
	\EndFunction
\end{algorithmic}
\end{algorithm}

\para{Algorithm}
The pseudocode of \mapper{} is given as Algorithm~\ref{alg:membership-query}.
It receives as input an initial cache content $\cache{\cacheContent_0}$, a policy trace $\policy{t} \in \finiteSequences{(\inputPolicy \times \outputPolicy)}$, and the cache semantics $\cacheSemantics{\cache{C}}$;
and it outputs $\mathit{true}$ if $\policy{t}$ belongs to the policy semantics $\policySemantics{\policy{P}}$, and $\mathit{false}$ otherwise.

$\mapper{}$ relies on the following  helper functions: %
\begin{asparaitem}
\item \(\mathit{probeCache}\) which, given a trace of blocks $\cache{q}$ and the cache semantics $\cacheSemantics{\cache{C}}$, accesses all blocks in $\cache{q}$ and returns whether the last block produces $\CacheHit$ or $\CacheMiss$ according to the cache semantics $\cacheSemantics{\cache{C}}$.

\item \(\textit{mapInput}\) which, given a policy input $\policy{ip}$ and a cache content $\cacheContent$, maps $\policy{ip}$ to a memory block $\cache{b}$.
If $\policy{ip}$ is
$\PolicyHit{i}$, the function returns $\element{\cacheContent}{i}$.
Otherwise (i.e., $\policy{ip}$ is $\PolicyMiss$), it returns a block $\cache{b}$ not in $\cacheContent$.

\item \(\textit{mapOutput}\) which, given a cache output $\cache{oc}$, a trace of blocks  $\cache{q}$,  a cache content $\cacheContent$, and the cache semantics $\cacheSemantics{\cache{C}}$, maps $\cache{oc}$ to the line containing the block that is evicted.
If $\cache{oc}$ is $\CacheHit$, the function returns $\PolicyNoEvict$.
Otherwise (i.e., $\cache{oc}$ is $\CacheMiss$), it returns the line $\policy{i}$ where the evicted block was stored. %

\item \(\mathit{findEvicted}\) which, given a trace of blocks $\cache{q}$, a cache content $\cacheContent$, and the cache semantics $\cacheSemantics{\cache{C}}$, determines which line has been evicted by the last block in $\cache{q}$.
For that, the function probes the cache with block traces $\cache{q} \concat \element{\cacheContent}{1}$, $\ldots$, $\cache{q} \concat \element{\cacheContent}{n}$ and determines which block resulted in $\CacheMiss$, i.e., the line that has been evicted by the last block in $\cache{q}$.
\end{asparaitem}

We are now ready to describe $\mapper{}$ in detail:
For each pair $\PolicyIo{ip}{op} \in \policy{t}$, the algorithm maps the input policy symbol $\policy{ip}$ to a memory block $\cache{b}$ ($\textit{mapInput}$ call at line~\ref{lst:interpol:mapinput}).
Then, the algorithm probes the cache to determine the result $\cache{oc}$ of accessing $\cache{b}$.
Next, the cache output $\cache{oc}$ is mapped to an output policy symbol $\policy{op'}$ ($\textit{mapOutput}$ call at line~\ref{lst:interpol:mapoutput}).
If $\policy{op}$ does not match the computed $\policy{op'}$, the algorithm returns $\mathit{false}$; else, the algorithm moves to the next pair of input/output policy symbols, and returns $\mathit{true}$ when the sequence is entirely processed.

For this, \mapper{} keeps track  of the sequence of blocks processed so far (through the $\cache{ic}$ variable, which is updated after every cache miss in line~\ref{lst:interpol:update}) and the
sequence of blocks processed so far (through the $\cache{ic}$ variable, which is updated with every call to \(\textit{mapInput}\)).  $\cache{ic}$ is used to set the cache $\cache{C}$ into the correct state before accessing the new block $\element{\cache{ic}}{i}$  (line~\ref{lst:interpol:probecache}), and for identifying the cache line that was last evicted ($\mathit{findEvicted}$ function).

\para{Correctness}

Theorem~\ref{theorem:policy-extraction-is-correct} states that, given a cache $\cache{C}$ with unknown policy $\policy{P}$, \mapper{} provides a sound, complete and terminating oracle for $\policy{P}$'s trace semantics.

\begin{restatable}{theorem}{soundnessOfCacheAbstraction}
\label{theorem:policy-extraction-is-correct}
Let $\cache{C}$ be a cache of associativity $n$  with initial content $\cache{\cacheContent_0} \in \CacheContents{n}$ and  policy $\policy{P}$.
Then, $\policySemantics{\policy{P}} = \{ \policy{t} \in \finiteSequences{(\inputPolicy \times \outputPolicy)} \mid \mapper{}(\cache{\cacheContent_0},\policy{t}, \cacheSemantics{\cache{C}}) = \mathit{true}   \}$.
\end{restatable}

As a corollary to Theorem~\ref{theorem:policy-extraction-is-correct}, we obtain a language-theoretic relationship between policies and caches.
Proposition~\ref{proposition:determinacy} states that, once we fix the initial cache content $\cache{\cacheContent_0}$ and associativity $n$, a replacement policy $\policy{P}$ uniquely determines the corresponding cache $\Cache{\policy{P}}{\cache{\cacheContent_0}}{n}$.

\begin{restatable}{proposition}{determinacy}
\label{proposition:determinacy}
Given two policies $\policy{P}$ and $\policy{P'}$ of associativity $n \in \Nat$ and an initial cache content $\cache{\cacheContent_0} \in \CacheContents{n}$,  then $\policySemantics{\policy{P}} = \policySemantics{\policy{P'}}$ iff $\cacheSemantics{\Cache{\policy{P}}{\cache{\cacheContent_0}}{n}} = \cacheSemantics{\Cache{\policy{P'}}{\cache{\cacheContent_0}}{n}}$.
\end{restatable}

Concretely, Proposition~\ref{proposition:determinacy} provides a theoretical justification for learning only the policy, since knowing the policy is equivalent to knowing the cache behavior.

\subsection{Equivalence Queries}\label{sec:learning:equivalence}

Equivalence queries between two Mealy machines \(M\) and \(M'\) are commonly implemented using {\em conformance testing}, which relies on a test suite (TS) of membership queries: If there is a membership query in TS on which \(M\) and \(M'\) disagree, the machines are clearly not equivalent. However, there are Mealy machines \(M\) for which there is no finite test suite that demonstrates non-equivalence for all machines \(M'\) with \(\semantics{M}\neq\semantics{M'}\). That is, the approximation of equivalence queries using finite membership tests is not complete~\cite{moermanNominalTechniquesBlack2019}.

In our approach, we hence aim for a weaker notion of completeness due to~\cite{moermanNominalTechniquesBlack2019}.
Namely, for a parameter \(m\) we say that a test suite TS is \emph{\(m\)-complete} for a hypothesized policy \(\policy{H}\), if there is no policy \(\policy{P}\) with less than $m$ control states and \(\policySemantics{\policy{H}}\neq\policySemantics{\policy{P}}\), such that both \(\policy{H}\) and \(\policy{P}\) agree on the TS.
With this, one can obtain the following guarantees for the equivalence test.\looseness=-1

\begin{theorem}\label{thm:eqtest}
	Let \(\policy{P}\) be an unknown replacement policy, \(\policy{H}\) a hypothesized policy, and TS an \(m\)-complete test suite for \(\policy{H}\) for some \(m \in \Nat\).
	If \(\policy{P}\) and \(\policy{H}\) agree on all queries of TS then either \(\policySemantics{\policy{H}}=\policySemantics{\policy{P}}\) or \(\policy{P}\) has more than \(m\) states.
\end{theorem}
We use existing algorithms, e.g. the Wp-Method~\cite{Khendek91testselection}, to compute $m$-complete suites for our hypothesized policy $\policy{H}$.

\subsection{Tool Implementation}\label{sec:learning:implementation}

We implement \mapper{} in a Java prototype on top of the LearnLib automata framework v0.14.0~\cite{Steffen2011}. This allows us to leverage state-of-the-art automata learning algorithms.

To access the cache semantics $\cacheSemantics{\cache{C}}$, our tool interacts either with a software-simulated cache or with \cachequery{} when targeting real hardware. %

Concretely:
\begin{asparaitem}
\item for the membership oracle, we implement \mapper{}, as described in \S~\ref{sec:learning:membership}.
\item for the equivalence oracle, we rely on the Wp-Method~\cite{Khendek91testselection} for computing test suites for conformance testing, as described in \S~\ref{sec:learning:equivalence}. Ideally, one would use a $m$-complete TS for $\policy{H}$, for $m$ as large as possible. Unfortunately the cost of computing $m$-complete test suite grows exponentially with $m$.
To achieve a good trade-off between completeness guarantees and complexity, we rely on $(\sizeof{\policy{H}}+k)$-complete TS, for a small constant $k$ which we call the {\em depth} of the suite.
\end{asparaitem}

Our main learning loop uses the $k$-deep conformance test for finding counterexamples. If the test fails, i.e., a counterexample for the current hypothesis is found, we refine the hypothesis and learning continues. Otherwise, the learning terminates and we output the current hypothesis.
At the end, we get the following overall guarantees.
\begin{corollary}\label{corollary:learning}
If our learning approach with \mapper{}, applied to a cache $\Cache{\policy{P}}{\cache{\cacheContent_0}}{n}$, returns a policy  $\policy{P'}$, then $\policySemantics{\policy{P}}=\policySemantics{\policy{P'}}$, or $\policy{P}$ has more than
$\sizeof{\policy{P'}}+k$ states.
\end{corollary}

To avoid cumbersome notation, from this point on we use \mapper{} to refer to both our algorithm and to our prototype tool integrating the algorithm with the learning loop.

%

%

%

%

%
%
%
%
%
%
%

%

%
%
%
 % end input ./learning.tex
 %
%
% start input ./tool.tex
%
\section{\cachequery: An Interface to Hardware Memory Caches}\label{sec:tool}

In this section we present \cachequery{}, a tool for querying silicon CPU caches.
\cachequery{} exposes an abstract interface to individual cache sets, and it frees the user from low-level details like slicing, mapping, virtual-to-physical translation, and profiling.
Concretely, \cachequery{} provides direct access to the trace semantics of hardware caches.

We first describe MemBlockLang, the domain-specific language for specifying inputs to \cachequery{}; then we describe \cachequery{}'s architecture and discuss some of its implementation challenges.

\subsection{Domain Specific Language}

We design {\em MemBlockLang} (MBL), a language that facilitates the writing of queries to caches.

A {\em query} is a sequence of one or more {\em memory operations}.
Each memory operation is specified as a block from a finite, ordered set of blocks $\CacheBlocks$, and it is decorated with an optional {\em tag} from \{\verb+?,!+\}.
The tag `\verb+?+' indicates that the access to the block should be profiled~\cite{intelbenchmark} to determine whether it results in a cache hit or miss; the tag `\verb+!+' indicates that the block should be invalidated (e.g., via \verb!clflush!); and no tag means that the block should just be accessed.

MBL features several macros that facilitate writing common query sequences:
\begin{asparaitem}

\item An {\em expansion} macro `\verb!@!' that produces a sequence of associativity many different blocks in increasing order, starting from the first element. For example, for associativity 8, \verb!@! expands to the sequence of blocks \verb!A B C D E F G H!.

\item A {\em wildcard} macro `\verb!_!' produces associativity many different queries, each one consisting of  a different block.
As for `\verb!@!', blocks are chosen in alphabetical order. %
For example, for associativity 8, \verb!_! expands to the set of queries \{\verb!A!, \verb!B!, \verb!C!, \verb!D!, \verb!E!, \verb!F!, \verb!G!, \verb!H!\}.

\item A {\em concatenation} macro $q_1 \circ q_2$ that concatenates each query in $q_1$'s expansion with each query in $q_2$'s expansion.
For instance, \verb!(A B C D)! $\circ$ \verb!(E F)! expands to  the query \verb!A B C D E F!.

\item An {\em extension} macro $q_1\left[q_2\right]$ that takes as input queries $q_1$ and $q_2$ and then creates $\sizeof{q_2}$ many copies of $q_1$ and extends each of them with a different element of $q_2$.
	For example, \verb!(A B C D)[E F]! expands to the set of queries \{\verb!A B C D E!, \verb!A B C D F!\}.

\item A power operator $(q)n$ that repeats a query macro $q$ for $n$ times. For example, \verb!(A B C)3! expands to the query \verb!A B C A B C A B C!.
\item A tag over ($q$) or [$q$] applies to every block in $q$. For example, \verb!(A B)?! expands to \verb!A? B?!.
\end{asparaitem}

MBL expressions can be given a formal semantics in terms of sets of queries
(cf.~Appendix~\ref{appendix:mbl:syntax-and-semantics}). %
For the purpose of presentation we omit such a formalization and focus on examples.

\begin{example}
For associativity 4, the query `\verb!@ X _?!' expands to `\verb!(A B C D)! $\circ$ \verb!X! $\circ$ \verb![A B C D]?!' or, equivalently, to the set of queries \{\verb!A B C D X A?!, \verb!A B C D X B?!, \verb!A B C D X C?!, \verb!A B C D X D?!\}.
This query performs an initial insertion (i.e., fills the cache with blocks \verb!A B C D!), accesses a block \verb!X! not in the cache, and probes all blocks \verb!A!, \verb!B!,  \verb!C!, and \verb!D! to determine which one has been replaced after the cache miss caused by \verb!X!.
This query implements the function \(\mathit{findEvicted}\) in Algorithm~\ref{alg:membership-query}.
\end{example}

\subsection{Architecture}

\cachequery~is split into two parts, described next. The backend is implemented in C as a Linux Kernel Module, while the frontend is implemented in Python 3.

\para{Frontend}
\cachequery's frontend expands MBL expressions into sets of queries.
The frontend provides two different execution modes: interactive and batch.
\begin{asparaitem}
	\item The {\em interactive} mode provides a REPL shell for executing queries, modifying configuration options, and dynamically choosing the target cache level and set. We use this mode as an interface for the learning algorithm
	\item The {\em batch} mode allows to run groups of predefined queries against different cache sets. This becomes useful for running batteries of tests, which, for instance, allows us to identify fixed leader sets (cf.~Appendix~\ref{appendix:leaders}). %
\end{asparaitem}

Furthermore, the frontend  uses LevelDB\footnote{LevelDB is a fast string key-value storage library: \url{https://github.com/google/leveldb}.} to cache query responses.
This improves performance by avoiding repeatedly issuing the same queries to the backend. %

\para{Backend}
\cachequery{}'s backend translates arbitrary que\-ries into sequences of memory accesses, generates the appropriate machine code with the corresponding profiling, executes the code in a low-noise environment, and finally returns traces of hits and misses.
We remark that profiling happens at the granularity of individual memory accesses, and it supports performance counters, time stamp counter, and counting core cycles, as demanded by the user.

The backend is implemented as a Loadable Kernel Module (LKM). This allows us to use APIs that provide fine-grained control over operating system details like virtual to physical address translation, interrupts, and preemption.

\begin{figure}[h]
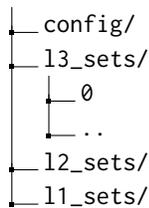

\begin{center}
\renewcommand*{\DTstyle}[1]{\texttt{#1}\hfill} %
\begin{minipage}{2cm}
\dirtree{%
.0 .
.1 config/.
.1 l3\_sets/.
.2 0.
.2 ...
.1 l2\_sets/.
.1 l1\_sets/.
}
\end{minipage}
\end{center}
\caption{LKM's virtual file system used by \cachequery{}}\label{fig:lkmDirectory}
\end{figure}

On load, the LKM allocates several pools of memory (one per cache level) and maps each memory block into its corresponding cache sets, one per cache level. This facilitates the address selection during code generation.

The backend provides a virtual file system interface, depicted in Figure~\ref{fig:lkmDirectory}, and all the communication is handled through read and write operations over virtual files. %
Specifically, writing query sequences into the cache set virtual file triggers the address selection and code generation, whereas reading from the cache set virtual file executes the generated code, and returns the sequence of hits and misses.

\subsection{Implementation Challenges}\label{sec:cachequery:challenges}
In the following we discuss some of the challenges in implementing \cachequery{}.

\para{Set Mapping}
The first challenge is identifying which memory addresses are mapped into which cache sets, i.e. which addresses are {\em congruent}.
For this, we need to know the number of cache sets, if these sets are virtually or physically indexed, and how the mapping is performed, i.e., which bits of the address are used for the mapping.
For most architectures, this information is publicly available~\cite{Irazoqui15systematicreverse,MauriceSNHF15}.
Otherwise, it is possible to infer it~\cite{cispa596} or to dynamically find congruent addresses~\cite{eviction19}.
\cachequery{} is completely parametric on the set mapping details, which need  only to be determined once per microarchitecture. %

\para{Cache Filtering}
When running queries against a low-level cache, say L3, one needs to make sure that the corresponding memory accesses do not hit higher-level caches such as L1.
To this end, \cachequery{} automatically evicts every accessed block from higher-level caches.
For instance, after accessing a block \verb!b! in L3, \cachequery{} automatically accesses non-interfering eviction sets (i.e. addresses that are congruent with \verb!b! in L2 and L1, but {\em not} congruent in L3) to ensure \verb!b!'s eviction from L2 and L1.

\para{Code Generation}
MBL expressions are first expanded into sets of queries, which are then dynamically translated into native code for execution.
Each query is implemented as a function returning a 64-bit mask with the hit/miss pattern of the profiled memory blocks.

The generated code includes the necessary profiling instructions (e.g., \texttt{rdtsc}), the conditional moves to update the output bit mask, and the additional memory loads to evict higher cache levels.
To support arbitrary queries, we use immediate load operations\footnote{That is, \texttt{movabs rax, qword [address]} (or in binary \texttt{0x48 0xa1 <imm>}).} serialized with memory fences, rather than the more common pointer chasing technique~\cite{Tromer2010}.

\para{Interferences}
To minimize noise during memory interactions, \cachequery{}  temporarily disables hardware prefetchers, hyper-threading, frequency scaling, and other cores.
To minimize cache interferences, we repeatedly allocate the generated code until it does not conflict with the target cache set.
Furthermore, the generated code is executed multiple times to reduce measurement noise.\looseness=-1

\subsection{Limitations}

Currently, \cachequery{} only supports data caches (not instruction caches) in Intel CPUs. While several parts of our implementation are architecture-agnostic, adding support for other architectures, such as AMD and ARM, will require manual effort.
Specifically, one would have to identify the (possibly undocumented) microarchitectural components whose state might affect the interaction with the cache, and manually disable or reset them between queries.

Likewise,
\cachequery{} currently runs on top of a fully-fledged Linux kernel. While facilitating development, this adds unnecessary complexity and non-determinism. Using a custom unikernel could provide a better suited environment for our experiments.

%

%

%
 % end input ./tool.tex
 %
%
% start input ./synthesis.tex
%
\section{Explaining Policies}\label{sec:program-synthesis}

In this section, we present our  approach for synthesizing explanations of replacement policies in the form of high-level programs, starting from our automata models.%

\para{Policy explanations}
We explain replacement policies in terms of four simpler rules:
\begin{inparaenum}[(a)]
\item a {\it promotion rule} describing how the control state is updated whenever there is a cache hit,
\item an {\it eviction rule} describing how to select the cache line to evict,
\item an {\it insertion rule} describing how the control state is updated whenever there is a cache miss, and
\item a {\it normalization rule} describing how to normalize the control state before or after a hit or a miss\footnote{Normalization is used in some policies to preserve control state invariants. For example, MRU updates the control state after a hit if all lines have age $0$.\looseness=-1}.
\end{inparaenum}
We borrow these terms from policy proposals from the hardware community~\cite{Jaleel2010}.

\para{Explanation template}
The main component of our synthesis approach is a program-level template for explanations, which is defined in terms of promotion, eviction, insertion, and normalization rules:
\begin{lstlisting}[style=CStyle, mathescape]
hit(state,line)::States$\times$Lines$\to$States
	state = promote(state,line)
	state = normalize(state,line)
	return state

miss(state)::States$\to$States$\times$Lines
	Lines idx = -1
	state = normalize(state,idx)
	idx = evict(state)
	state[idx] = insert(state,idx)
	state = normalize(state,idx)
	return $\langle$state, idx$\rangle$
\end{lstlisting}
The template models control states as arrays mapping cache lines to their so-called ages.
The concrete age values (of type \inlineCode{Nat}) are left as holes to be instantiated during the synthesis.
Additionally, the template consists of two functions:

The function \inlineCode{hit} describes how the control state is updated whenever there is a cache hit.
The function takes as input a control state \inlineCode{state} and a cache line \inlineCode{line}, updates the control state using the promotion rule, normalizes it, and returns the new state.

In contrast, the function \inlineCode{miss} modifies the control state in case of a cache miss.
The function takes as input a control state \inlineCode{state}, normalizes it, detects the cache line \inlineCode{idx} to evict using the eviction rule, updates the age of the evicted line using the insertion rule, and finally normalizes again the ages.\looseness=-1

We remark that our templates---with promotion, eviction, insertion, and normalization rules---formalize well-known concepts and building blocks used by cache designers~\cite{Jaleel2010}.

\para{Generators}
Our template specifies several \textit{generators} for the rules.
Generators are programs with holes that can be instantiated during synthesis.
Each of the holes can be instantiated with expressions generated from specific grammars, which constraint the synthesis' search space.
To illustrate, this is a generator for the promotion rule: %
\begin{lstlisting}[style=CStyle, mathescape]
promote(state,pos)::States$\times$Lines$\to$States
	States final = state
	if(??{$\mathit{boolExpr}($state[pos]$)$}) // Update line
		final[pos] = ??{$\mathit{natExpr}($state[pos]$)$}
  	for(i in Lines) // Update rest
  		if(i $\neq$ pos $\wedge$ ??{$\mathit{boolExpr}($state[pos]$,$state[i]$)$})
  			final[i] = ??{$\mathit{natExpr}($state[i]$)$}
  	return final
\end{lstlisting}
The generator takes as input a control state and a cache line and returns the updated control state.
The updated state is derived by first conditionally modifying the age of the accessed line and later iterating over the remaining cache line and conditionally updating them.

All conditions and update expressions are encoded as holes that refer to template variables.
For instance, the hole \inlineCode{??\{}$\mathit{boolExpr}($\inlineCode{state[pos],state[i]}$)$\inlineCode{\}} can be instantiated with a conjunction of equalities and inequalities that refer to natural numbers and to \inlineCode{state[pos]} and \inlineCode{state[i]}, whereas  \inlineCode{??\{}$\mathit{natExpr}($\inlineCode{state[i]}$)$\inlineCode{\}} can be instantiated with an arbitrary sequence of additions and subtractions that refer to natural numbers and to  \inlineCode{state[i]}.

In general, our grammar generators can refer to constants, line indices (like \inlineCode{pos} and \inlineCode{i} in the \inlineCode{promote} example), and ages (like \inlineCode{state[pos]} in the \inlineCode{promote} example).
We also implement a simplified version of our generators that (1) fix the \inlineCode{normalize} rule to the identity function, and (2) restrict the grammar to only refer to constants and ages.

\para{Constraints}
Given a policy $\policy{P}$, we construct a formula $\varphi_{\policy{P}}$ encoding $\policy{P}$'s transition relation $\evalPol{}$ in terms of our template's \inlineCode{hit} and \inlineCode{miss} functions.
In our encoding, we associate $\policy{P}$'s control states with logical variables.
Concretely, we map each control state $\policy{\controlState_i}$ in $\policy{P}$ to a corresponding variable $\text{\inlineCode{cs}}_i$. %
The constraint  $\varphi_\policy{P}$ is defined as follows, where $\policy{\controlState_1}, \ldots, \policy{\controlState_m}$ are all $P$'s control states:
\begin{align*}
&\exists \text{\inlineCode{cs}}_1, \ldots, \text{\inlineCode{cs}}_{m}.\ \bigwedge_{1 \leq i,j \leq |P| \wedge i \neq j} \text{\inlineCode{cs}}_i \neq \text{\inlineCode{cs}}_j \wedge \\
& \quad \bigwedge_{\PolicyTransitionFunction(\policy{\controlState_k},\PolicyHit{i}) = \policy{\controlState_l}} \text{\inlineCode{hit}}(\text{\inlineCode{cs}}_k, i) = \text{\inlineCode{cs}}_l\ \wedge \\
& \quad\bigwedge_{
\substack{\PolicyTransitionFunction(\policy{\controlState_k},\PolicyMiss) = \policy{\controlState_l} \wedge \\\PolicyOutputFunction(\policy{\controlState_k},\PolicyMiss) = \PolicyEvict{i}}} \text{\inlineCode{miss}}( \text{\inlineCode{cs}}_k) = \tup{ \text{\inlineCode{cs}}_l,i}
\end{align*}
The existential quantification and the first conjunct ensure that there are $m$ concrete control states (one per control state in $\policy{P}$).
The second and third conjuncts ensure that the \inlineCode{hit} and \inlineCode{miss} functions behave as specified by $\policy{P}$.

\para{Synthesis}
To synthesize an explanation for a learned policy $\policy{P}$, we query a syntax-guided synthesis solver for an instance of our  template that satisfies the constraint $\varphi_\policy{P}$.
The solver, then, either returns a program $\mathit{Prg}$ that instantiates the holes in the template in a way that satisfy $\varphi_\policy{P}$, or terminates without finding a model (in case our template cannot represent $\policy{P}$).

\begin{example}
Let $\policy{P}$ be the LRU policy  with associativity 2 given in Example~\ref{example:lru-policy}.
The constraint $\varphi_\policy{P}$ is as follows:
\begin{multline*}
	\exists\ \text{\inlineCode{cs}}_0, \text{\inlineCode{cs}}_1.\ \text{\inlineCode{cs}}_0 \neq \text{\inlineCode{cs}}_1 \wedge  \text{\inlineCode{hit}}(\text{\inlineCode{cs}}_0, 1) = \text{\inlineCode{cs}}_0 \wedge\\
\text{\inlineCode{hit}}(\text{\inlineCode{cs}}_0, 0) =\text{\inlineCode{cs}}_1 \wedge
  \text{\inlineCode{hit}}(\text{\inlineCode{cs}}_1, 0) = \text{\inlineCode{cs}}_1 \wedge \text{\inlineCode{hit}}(\text{\inlineCode{cs}}_1, 1) = \text{\inlineCode{cs}}_0 \wedge \\
 \text{\inlineCode{miss}}(\text{\inlineCode{cs}}_0) = \tup{ \text{\inlineCode{cs}}_1, 0  } \wedge \text{\inlineCode{miss}}(\text{\inlineCode{cs}}_1) = \tup{ \text{\inlineCode{cs}}_0, 1  }\enspace .
\end{multline*}
\end{example}

Whenever the solver synthesizes a program that satisfy the given constraints, we can lift the correctness guarantee of our approach also to the synthesized program $\mathit{Prg}$.
Indeed, the solver's soundness, the template's determinism, and the constraint $\varphi_\policy{P}$ ensure that $\mathit{Prg}$ behaves exactly as the learned policy $\policy{P}$ on the concrete control states.

\para{Limitations}
Although our templates support a large class of policies (see \S~\ref{sec:evaluation:explanations}), they cannot explain arbitrary  policies.
For instance, we model control states by associating an age to each cache line.
Hence, policies with a global control state, such as PLRU, are not supported.
Similarly, policies that do not follow the structure of the promotion, eviction, insertion, and normalization rules are not supported.

%
%
%
%
%
%
 % end input ./synthesis.tex
 %
%
% start input ./evaluation.tex
%

%

%

\section{Case Study: Learning from Software-Simulated Caches}\label{sec:evaluation:simulators}

This section reports on a case study where we use \tool{} to learn well-known replacement policies from software-simulated caches implementing such policies.

This case study's goals is to evaluate \tool{}'s efficiency and scalability across different classes of replacement policies, without the overhead introduced by interacting with real hardware.
This case study also provides a basis for comparing \tool{} with prior approaches~\cite{grueda13,cispa596}.

\para{Setup}
We implemented software-simulated caches (parametric in the cache's associativity) for 7 commonly used replacement policies:
First In First Out (FIFO), Least Recently  Used (LRU), Pseudo-LRU (PLRU)~\cite{Handy1993}, Most Recently Used (MRU)~\cite{mru1992}, LRU Insertion Policy (LIP)~\cite{Qureshi2007}, and HP and FP variants of Static Re-reference Interval Prediction (SRRIP)~\cite{Jaleel2010} with 4 ages.
We simulate the policies with associativity ranging from 2 to 16.\footnote{Some policies constraint the possible associativities. For instance, PLRU policies are well-defined only for associativities that are powers of 2.}
For each policy and associativity, we use \tool{} to learn the policy with a timeout of $36$ hours.
We record the time needed to learn the automaton and the learned automaton's number of states.
In our experiments, we set the test suite depth $k$ to $1$ (\S~\ref{sec:learning:implementation}), which proves sufficient for discovering counterexamples.

%
%

%

%
%
%

\iffalse
\begin{figure}[ht]
\begin{subfigure}[b]{\columnwidth}
\hfill
% inside_import 
% before 
% ignored 
% args width=\columnwidth
% full_filename figs/simul-states-assoc
% after 
\includegraphics[width=\columnwidth]{figs/simul-states-assoc}
\caption{Number of states in learned automata.}\label{figure:simulators:states}
\end{subfigure}
\begin{subfigure}[b]{\columnwidth}
%
% inside_import 
% before 
% ignored 
% args width=\columnwidth
% full_filename figs/simul-time-states
% after 
\includegraphics[width=\columnwidth]{figs/simul-time-states}
\caption{Learning time.}\label{figure:simulators:time}
\end{subfigure}
\caption{
Learning policies from software-simulated caches. Timeout after 36 hours of computation.
}\label{figure:simulators}
\end{figure*}
\comment[pv]{Figure~\ref{figure:simulators:time} tests different representation.}
\fi

%
\begin{table}%
\centering
\small
\caption{
Learning policies from software-simulated caches (with  36 hours timeout). We omit FIFO's intermediate results. %
}\label{table:simulators}
\begin{tabular}{c r r r}
\toprule
\textbf{Policy} & \textbf{Assoc.} & \textbf{\# States} & \textbf{Time} \\
\midrule[0.08em]
\multirow{3}{*}{\textit{FIFO}} & 2 & $2$ & $0$\,h $0$\,m $0.14$\,s \\
& $\ldots$ & $\ldots$ & $\ldots$ \\
& 16 & $16$ & $0$\,h $0$\,m  $0.38$\,s \\
\midrule[0.01em]
\multirow{4}{*}{\textit{LRU}} & 2 & $2$ & $0$\,h $0$\,m $0.10$\,s \\
& 4 & $24$ & $0$\,h $0$\,m $0.22$\,s \\
& 6 & $720$ & $0$\,h $0$\,m $32.70$\,s \\
\midrule[0.01em]
\multirow{4}{*}{\textit{PLRU}} & 2 & $2$ & $0.10$\,s \\
& 4 & $8$ & $0.22$\,s \\
& 8 & $128$ & $1.46$\,s \\
& 16 & $32768$ & $34$\,h $18$\,m $25$\,s \\ %
\midrule[0.01em]
\multirow{6}{*}{\textit{MRU}} & 2 & $2$ & $0$\,h $0$\,m $0.10$\,s \\
& 4 & $14$ & $0$\,h $0$\,m $0.16$\,s \\
& 6 & $62$ & $0$\,h $0$\,m $0.61$\,s \\
& 8 & $254$ & $0$\,h $0$\,m $8.82$\,s \\
& 10 & $1022$ & $0$\,h $5$\,m $58$\,s \\ %
& 12 & $4094$ & $3$\,h $59$\,m $20$\,s \\ %
\midrule[0.01em]
\multirow{3}{*}{\textit{LIP}} & 2 & $2$ & $0$\,h $0$\,m $0.10$\,s \\
& 4 & $24$ & $0$\,h $0$\,m $0.26$\,s \\
& 6 & $720$ & $0$\,h $0$\,m $31.97$\,s \\
\midrule[0.01em]
\multirow{3}{*}{\textit{SRRIP-HP}} & 2 & $12$ & $0$\,h $0$\,m $0.16$\,s \\
& 4 & $178$ & $0$\,h $0$\,m $1.46$\,s \\
& 6 & $2762$ & $0$\,h $9$\,m $38$\,s \\ %
\midrule[0.01em]
\multirow{3}{*}{\textit{SRRIP-FP}} & 2 & $16$ & $0$\,h $0$\,m $0.19$\,s \\
& 4 & $256$ & $0$\,h $0$\,m $7.27$\,s \\
& 6 & $4096$ & $2$\,h $30$\,m $51$\,s \\ %
\bottomrule
\end{tabular}
\end{table}

\para{Results}
Table~\ref{table:simulators} reports the time taken by \tool{} to learn the policies and the number of states of the resulting automata.
We highlight the following:
\begin{asparaitem}
\item Except for FIFO, the learning time grows roughly exponentially with associativity. \tool{} learns FIFO and PLRU up to associativity 16.\looseness=-1
\item Prior approaches for permutation-based policies~\cite{cispa596} can learn only FIFO, LRU, and PLRU from our experimental setup.
In contrast, \tool{} learns policies such like MRU, LIP, SRRIP-HP, and SRRIP-FP (up to associativities 12, 6, 6, and 6 respectively).

\item Prior general purpose approaches~\cite{grueda13} learn MRU only up to associativity 5 and timeout after 72 hours for larger associativities.
	In contrast, \tool{} learns MRU up to associativity 12 and takes $600$\,milliseconds for associativity 6. %

\end{asparaitem}

Alternative approaches exist that leverage different heuristics, like random walks, for a deeper counterexample exploration. These approaches generally enable faster hypothesis refinement, and hence better performance. However, we opted for a default and deterministic setup, and leave a more thorough performance evaluation for future work.

\para{Platform}
We run all experiments on a Linux virtual machine (kernel 4.9.0-8-amd64) with Debian 9.0, Java OpenJDK v1.8.0\_222, running on a Xeon Gold 6154 CPU (with 72 virtual cores), and 64 GB of RAM.
We execute the experiments
in parallel using a single virtual core for each policy. %

\section{Case Study: Learning from Hardware}\label{sec:evaluation:hardware}
In this section we report on a case study where we use \tool{} and \cachequery{} to learn policies from real hardware.
The case study's goals are
(1) to determine whether \tool{} can learn policies directly from hardware  using \cachequery{} as an interface,
and (2) to understand the additional challenges involved with learning policies from hardware.

\begin{table}[h]
\centering
\small
\caption{Processors' specifications~\cite{ia2018,Irazoqui15systematicreverse,MauriceSNHF15}.}
\label{table:machinespecs}
\begin{tabular}{c r r r r}
\toprule
\textbf{CPU} & \textbf{Cache level} & \textbf{Assoc.} & \textbf{Slices} & \textbf{Sets per slice} \\
\midrule[0.08em]
\multirow{3}{*}{
\shortstack[c]{
\textit{i7-4790}\\
\textit{(Haswell)}
}
} & L1 & $8$ & $1$ & $64$ \\
& L2 & $8$ & $1$ & $512$ \\
& L3 & $16$ & $4$ & $2048$ \\
\midrule[0.01em]
\multirow{3}{*}{
\shortstack[c]{
\textit{i5-6500}\\
\textit{(Skylake)}
}
} & L1 & $8$ & $1$ & $64$ \\
& L2 & $4$ & $1$ & $1024$ \\
& L3 & $12$ & $8$ & $1024$ \\
\midrule[0.01em]
\multirow{3}{*}{
\shortstack[c]{
\textit{i7-8550U}\\
\textit{(Kaby Lake)}
}
} & L1 & $8$ & $1$ & $64$ \\
& L2 & $4$ & $1$ & $1024$ \\
& L3 & $16$ & $8$ & $1024$ \\
\bottomrule
\end{tabular}
\end{table}

\begin{table*}[ht!]
\centering
\small
\caption{Results of learning policies from hardware caches. $\dagger$ indicates that the associativity has been virtually reduced using CAT. The `Sets' column specifies the analyzed cache sets (unless otherwise specified, the findings apply to all slices). F+R denotes the use of Flush+Refill to reset the cache set state.}
\label{table:learning-from-hardware}
\begin{tabularx}{\textwidth}{ c c c X c c c}
\toprule
\textbf{CPU} & \textbf{Level} & \textbf{Assoc.} & \textbf{Sets} & \textbf{States} & \textbf{Policy} & \textbf{Reset Seq.} \\ %
\midrule[0.01em]
\multirow{4}{*}{
\shortstack[c]{
\textit{i7-4790}\\
\textit{(Haswell)}
}} & L1 & $8$ & $0-63$ & $128$ & PLRU & \texttt{@ @} \\ %
 & L2 & $8$ & $0-511$ & $128$ & PLRU & F+R \\ %
 & \multirow{2}{*}{L3} & \multirow{2}{*}{$16$} & $512-575$ {\em (only for slice $0$)} & -- & -- & -- \\ %
& & & $768-831$ \em{(only for slice $0$)} & -- & -- & -- \\ %
\midrule[0.01em]
\multirow{3}{*}{
\shortstack[c]{
\textit{i5-6500}\\
\textit{(Skylake)}
}} & L1 & $8$ & $0-63$ & $128$ & PLRU & F+R \\ %
 & L2 & $4$ & $0-1023$ & $160$ & {\em New1} & \texttt{D C B A @} \\ %
 & L3 & $4^\dagger$ & 0 33 132 165 264 297 396 429 528 561 660 693 792 825 924 957 & $175$ & {\em New2} & F+R \\ %
\midrule[0.01em]
\multirow{3}{*}{
\shortstack[c]{
\textit{i7-8550U}\\
\textit{(Kaby Lake)}
}
} & L1 & $8$ & $0-63$ & $128$ & PLRU & F+R \\ %
 & L2 & $4$ & $0-1023$ & $160$ & {\em New1} & \texttt{D C B A @} \\ %
 & L3 & $4^\dagger$ & 0 33 132 165 264 297 396 429 528 561 660 693 792 825 924 957 & $175$ & {\em New2} & F+R \\ %
\bottomrule
\end{tabularx}
\end{table*}

\subsection{Setup}

We analyze the  L1, L2, and L3 caches of the Intel i7-4790 (Haswell), i5-6500 (Skylake), and i7-8550U (Kaby Lake) processors (see Table~\ref{table:machinespecs} for the specifications) using \tool{} and \cachequery{}.
We encounter the following challenges:
\begin{asparaitem}
\item For some policies, \tool{} does not scale to the large associativities used in L3 caches.
To overcome this, we use Intel's CAT technology~\cite{intelrdt} to virtually reduce L3 associativity to 4 for the i5-6500 (Skylake) and i7-8550U (Kaby Lake) processors. CAT is not supported by i7-4790 (Haswell). %

\item Modern L3 caches often implement adaptive replacement policies~\cite{Qureshi2006,Qureshi2007, Jaleel2010}, where separate groups of {\em leader} cache sets implement distinct replacement policies and the remaining {\em follower} sets switch between these policies dynamically.
We \textit{only} learn leader sets' policies (cf.~Appendix~\ref{appendix:leaders}).
\item Our membership oracle $\mapper{}$ (Algorithm~\ref{alg:membership-query}) relies on the assumption that traces are executed from a fixed initial state.
In practice, this leads to a bootstrapping problem: knowing the reset sequence (i.e., a sequence of memory accesses that brings the cache into a fixed initial state) is a prerequisite for learning the policy, but computing the reset sequence requires knowledge about the policy itself.
On many CPUs, cache sets can be reset by Flush+Refill, i.e., by invalidating the entire content (with \texttt{clflush} or \texttt{wbinvd} instructions) and accessing associativity-many different blocks (with the `\texttt{@}' MBL macro).
For CPUs where this is not the case, we manually identify reset sequences for each cache (see Table~\ref{table:learning-from-hardware}).
This is enabled by the fact that {\em incorrect} reset sequences lead to nondeterministic behavior (equal prefixes produce different outputs), which triggers errors in the learning algorithm.
\end{asparaitem}

\para{Platform}
We run \cachequery{} on three different machines equipped with the three processors.
Additionally, we run \tool{} on the same platform described in \S~\ref{sec:evaluation:simulators}.
The communication between \tool{} and \cachequery{} happens over SSH in a local network.

\subsection{Results}

\para{Learned policies}
Table~\ref{table:learning-from-hardware} summarizes the learned policies.
We highlight the following findings:
\begin{asparaitem}
	\item For all processors' L1 caches and for Haswell's L2 cache, \tool{} learns the same policy, that is, a tree-based PLRU policy.
	We identified this policy by checking its equivalence to a manually implemented PLRU automaton.
	This result confirms common folklore that Intel processors implement PLRU in their L1 policy.
	\item For Skylake's and Kaby Lake's L2 caches, \tool{} learns a previously undocumented policy.
	This policy is indicated as \textit{New1} in Table~\ref{table:learning-from-hardware}, and we further discuss it in \S~\ref{sec:evaluation:explanations}.
	\item For Skylake's and Kaby Lake's L3 caches, \tool{} learns a previously undocumented policy for the leader sets.
	This policy is indicated as \textit{New2} in Table~\ref{table:learning-from-hardware}, and we further discuss it in \S~\ref{sec:evaluation:explanations}.
	Additionally, we confirm the mapping of Skylake's leader sets~\cite{eviction19}, and we discover that Kaby Lake follows the same mapping.
	\item For Haswell's L3 cache, \tool{} cannot learn the replacement policy.
	This is due to (1) i7-4790 not supporting CAT, and (2) one of the leader sets showing a non-deterministic behavior.
\end{asparaitem}

\para{Cost of learning from hardware}
Learning policies from hardware caches comes with a significant overhead when compared with learning from software-simulated caches.
This is due to (1) communication overhead between \tool{} and \cachequery{}, and (2) \cachequery{} overhead for code generation and profiling.
We separately analyze the impact of (1) and (2).

For (1), we compare the time needed to learn a PLRU policy with associativity 8 from a software-simulated cache and from \cachequery{} where every MBL query  hits the LevelDB cache (i.e., the results of the MBL queries on the real hardware have been precomputed).
Learning from the software-simulated cache takes $1.46$\,s (cf.~Table~\ref{table:simulators}), while learning from \cachequery{} takes  $2247$\,s, resulting in a $1500$x overhead.

For (2), we measure the time taken to execute a single MBL query `\verb!@ M _?!' across cache levels.
The averaged query execution time (across 100 executions on the i5-6500 Skylake processor) is $16$\,ms on L1, $11$\,ms on L2, and $20$\,ms on L3.
We remark that learning the PLRU policy with associativity 8 requires more than 50'000 MBL queries.

\section{Case Study: Synthesizing Explanations}\label{sec:evaluation:explanations}

This section reports on a case study where we use the synthesis approach from \S~\ref{sec:program-synthesis} to derive policy explanations for the automata learned in \S\S~\ref{sec:evaluation:simulators}--\ref{sec:evaluation:hardware}.
This case study's goals are
(1) to evaluate if our approach can explain the replacement policies learned in \S\S~\ref{sec:evaluation:simulators}--\ref{sec:evaluation:hardware},
and (2) to determine whether the synthesized explanations can help in understanding previously undocumented policies.

\subsection{Setup}
We encoded our template (and all rules generators) from \S~\ref{sec:program-synthesis} in Sketch~\cite{sketch2009}.
We use Sketch to synthesize explanations for all the policies from \S~\ref{sec:evaluation:simulators} (i.e., FIFO, LRU, PLRU, MRU, LIP, SRRIP-HP, and SRRIP-IP) and for the undocumented policies \textit{New1} and \textit{New2} from \S~\ref{sec:evaluation:hardware}.
In our experiments, we fix the associativity to 4.

For all policies, we synthesize explanations using Sketch\footnote{
Sketch uses a random seed to explore the search space.
Hence, Sketch might synthesize different explanations that satisfy the constraints.
In our experiments, we fix the seed to \texttt{--slv-seed 1337}.
}, and record the time needed to synthesize an explanation that satisfies our constraints.
During synthesis, we bound both the size of natural numbers and the recursion depth of our grammar generators. For associativity 4, we choose a size bound of 4 and recursion depth bound of 2. We explore the synthesis space incrementally until we find a solution or the space is exhausted.
For each policy, we first try synthesizing an explanation using the simplified template from \S~\ref{sec:evaluation:simulators} (which we refer to as \textit{Simple} template), and if we cannot synthesize a solution, then we try using the more general template  from \S~\ref{sec:evaluation:simulators} (which we refer to as \textit{Extended} template).

\para{Platform}
We run the experiments on the same platform as in \S~\ref{sec:evaluation:simulators} and use Sketch v1.7.5, with a single thread.

\subsection{Results}

Table~\ref{table:sketches} summarizes the results of our synthesis approach.
We highlight the following findings:
\begin{asparaenum}
\item Our approach successfully explains the FIFO, LRU, and LIP policies using the \textit{Simple} template in less than $5$\,s.
\item Our approach synthesizes explanations for the MRU, SRRIP-HP, SRRIP-FP, $\mathit{New1}$, and $\mathit{New2}$ policies using the \textit{Extended} template.
The synthesis time varies (from {\textasciitilde{}}$40$\,s for MRU to {\textasciitilde{}}$4.5$\, days for SRRIP-HP), but it is roughly correlated with the number of states.

\item Our current templates do not encompass PLRU. The main reason is that PLRU uses a tree-based data structure as global control state, rather than a local, per-line control state as in our templates.
Supporting tree-based policies would require modifying our templates to handle  a global state and extending our grammars with operators for the traversal of tree-based structures.
Synthesis, however, did not successfully terminate in our initial experiments for this enhanced template.
Thus, we opted in favor of  simpler and more general templates that allow us to explain a broader set of policies in reasonable time, even at the cost of not supporting special tree-based global policies like PLRU.
\end{asparaenum}

\begin{table}[h]
\centering
\small
\caption{Synthesizing explanations for policies (of associativity 4).
In the \textit{Simple} template, \inlineCode{normalize} is fixed to the identity function and the grammar for expressions is simpler.
In contrast, the \textit{Extended} template supports the \inlineCode{normalize} rule and has a more expressive expression grammar.}
\label{table:sketches}
\begin{tabular}{c c r c r}
\toprule
\textbf{Policy} & \textbf{States} & \textbf{Template} & \textbf{Execution Time} \\
\midrule[0.08em]
FIFO & $4$ & {\em Simple} & $0$\,h $0$\,m $0.18$\,s \\
LRU & $24$ & {\em Simple} & $0$\,h $0$\,m $0.81$\,s\\
PLRU & $8$ & --- & --- \\
LIP & $24$ & {\em Simple} & $0$\,h $0$\,m $4.36$\,s \\
MRU & $14$ & {\em Extended} & $0$\,h $0$\,m $39.80$\,s \\
SRRIP-HP & $178$ & {\em Extended} & $105$\,h $28$\,m $30$\,s \\ %
SRRIP-FP & $256$ & {\em Extended} & $48$\,h $30$\,m $25$\,s \\ %
{\em New1} & $160$ & {\em Extended} & $9$\,h $36$\,m $9$\,s \\ %
{\em New2} & $175$ & {\em Extended} & $26$\,h $4$\,m $22$\,s \\ %
\bottomrule
\end{tabular}
\end{table}

\para{Explaining \textit{New1} and \textit{New2}}
Sketch successfully synthesize explanations for the previously undocumented policies \textit{New1} and \textit{New2} from \S~\ref{sec:evaluation:hardware}.
Below we provide a high-level description of the policies.
We include the complete synthesized programs in~Appendix~\ref{appendix:pseudocode}. %

The {\em New1} policy is defined by:
\begin{itemize}
	\item The initial control state is $\{3,3,3,0\}$.
	\item {\bf Promote:} Set the accessed line's age to $0$.
	\item {\bf Evict:} Select the first line, starting from left, whose age is $3$.
	\item {\bf Insert:} Set the evicted line's age to $1$.
	\item {\bf Normalize:} After a hit or a miss, while there is no line with age $3$, increase the age of all lines by $1$ except for the just accessed/evicted line.
\end{itemize}

The {\em New2} policy is defined by:
\begin{itemize}
	\item The initial control state is $\{3,3,3,3\}$.
	\item {\bf Promote:} If the accessed line has age $1$ set it to $0$, otherwise set it to $1$.
	\item {\bf Evict:} Select the first line, starting from left, whose age is $3$.
	\item {\bf Insert:} Set the evicted line's age to $1$.
	\item {\bf Normalize:} After a hit or miss, while there is no line with age $3$, increase all lines by $1$.
\end{itemize}

In contrast to the automata models, our high-level representation allow us to compare the previously undocumented policies with known ones.
Concretely, both {\em New1} and {\em New2} are variants of the SRRIP-HP policy, defined in~\cite{Jaleel2010}.
The main difference appears in the normalization rule, where  SRRIP-HP  normalizes the ages (by increasing all ages by $1$ while there is no line with age $3$) only before a miss.

%
%
%
%
%
%
%
%

%
%
%
 % end input ./evaluation.tex
 %
%
% start input ./discussion.tex
\section{Discussion}\label{sec:discussion}

\para{Threats to validity}

\begin{asparaitem}
\item \textit{Hardware interface:}
\cachequery{} employs several  mechanisms (\S~\ref{sec:cachequery:challenges}) to eliminate noise %
and to provide an interface to a cache whose state is determined by explicit memory loads. 
However, a replacement policy could take into account hints from hardware prefetchers, different cache levels, or other partitions created by CAT~\cite{dawg2018}. We suppress these effects by design and thus potentially learn a restricted state machine.

\item \textit{Automata learning:}
Our automata learning approach provides precise correctness guarantees (Corollary~\ref{corollary:learning}).
These guarantees rely on two assumptions: (1) the policy under learning is a deterministic finite state machine, and (2) the test suite of depth $k$ is large enough to find counterexamples during the learning.
Violating any of these assumptions may result in an incorrect policy.

\end{asparaitem}

\para{Scalability}
Our approach can successfully learn policies up to associativity 16 (for software simulators, see~\S~\ref{sec:evaluation:simulators}) and 8 (for hardware caches, see~\S~\ref{sec:evaluation:hardware}).
\mapper{} is key to make automata learning scale to this extent since it exploits data-independence symmetries to significantly reduce the learning state-space.
For many policies, however, scalability is still limited by the state-space's exponential growth with respect to associativity. %
Potential paths forward include identifying and exploiting further symmetries to reduce the state-space, learning abstractions rather than full models, or giving up on the correctness guarantees. %
Due to its modularity, our approach is well-suited for integrating such variants (and future improvements) in automata learning and program synthesis.\looseness=-1

%
%
%
%
%
%
%

%
%
%

%
%
%
%
%
%

 % end input ./discussion.tex
 %
%
% start input ./related.tex
%
\section{Related Work}\label{sec:related}

\para{Model learning}
For related work on automata learning techniques for black-box systems we refer the interested reader to Vaandrager's survey paper~\cite{vaandragerModelLearning2017}.

\para{Reverse-engineering cache policies}
Abel and Reineke~\cite{cispa596} design an efficient algorithm for learning permutation base replacement policies, a class of policies that include LRU, PLRU, and FIFO. They use an adhoc approach to reverse engineer two variants of PLRU that employ randomization~\cite{AbelISPASS214}.

Guillem Rueda's master thesis~\cite{grueda13} studies how register automata learning can serve to learn a broader class of replacement policies, in comparison to permutation based policies, including MRU. This is an interesting and novel approach, however, their method does not scale in practice (\S~\ref{sec:evaluation:simulators}).

Wong~\cite{ReplacementPolicy2013} notices the use of adaptive policies in Intel's Ivy Bridge, and tries to identify the new implemented policies guided by recent papers~\cite{Qureshi2006,Jaleel2010}, without complete success.

In concurrent work, Abel and Reineke extend nanoBench~\cite{nanobench2019, nanobenchcache2019} to reverse engineer cache replacement policies. In contrast to our approach, they proceed by producing random sequences and comparing the results from hardware against a pool of {\textasciitilde{}}300 software-simulated caches.
While this approach is less general and the results lack correctness guarantees, in practice, it proves highly efficient and accurate.
In fact, we are able to validate several of their findings.

\para{Security}
Rowhammer.js~\cite{rowhammerjs2016} tests thousands of eviction strategies, memory access patterns with high eviction rate, in order to identify efficient strategies to mount a rowhammer attack from a web browser.
Recent attacks~\cite{cacheaudit-tissec,dawg2018,lruleak2019} show how detailed knowledge about the replacement policy state can leak information, in contrast to the common content based leak.
Similarly, Briongos~et~al.~\cite{ReloadRefresh2019} exploit a new cache side-channel attack leveraging changes in the policy state to bypass mechanisms based on monitoring of cache misses. For this, they attempt to explain the behavior of the replacement policy on several modern Intel CPUs. While their description is not completely accurate, it is enough to prove their attack.

Detailed policy models, such as the ones we provide, enable one to systematically compute optimal eviction strategies, and to unveil new sophisticated cache attacks.

\para{Custom kernels}
Recent research projects have developed custom kernels and hypervisors for specialized tasks that require extremely high performance, precise measurements, or access to privileged modes.
These environments provide complete control over the hardware improving testing and reproducibility.
Some examples include angryOS~\cite{Koppe2017}, which has been used for reverse engineering microcode, or Sushi Roll~\cite{SushiRoll2019}, a highly deterministic kernel, initially designed for fuzzing, converted into a cycle-by-cycle CPU introspection tool.\looseness=-1

Implementing interfaces as \cachequery{} on a custom kernel can provide a better environment for high performance and predictability, ultimately enabling the use of learning methods for other undocumented microarchitectural components, like prefetchers, branch predictors, or data buffers.
 % end input ./related.tex
 %
%
% start input ./conclusions.tex
%

\section{Conclusions}\label{sec:conclusions}

We presented a practical end-to-end solution for learning hardware cache replacement policies. In our experiments we were successful in inferring human-readable descriptions of cache replacement policies used in recent Intel processors, including 2 previously undocumented policies.

Our approach relies on two contributions that enable us to tackle the problem using off-the-shelf techniques for automata learning and program synthesis:
(1) \cachequery{}, a tool that provides a clean interface for interacting with hardware memory caches, and (2) \mapper{}, an algorithm that provides a direct interface to the replacement policy by abstracting from the cache content.

Both our contributions are independent and ready to use in alternative workflows, such as advanced learning approaches~\cite{Cassel2018,nondeterministicmealy2014} or manual analysis.

\medskip
\section*{Acknowledgments}
We thank the anonymous reviewers and our shepherd Mangpo Phothilimthana  for insightful comments.
This work was supported by
a grant from Intel Corporation,
Atracci\'on de Talento Investigador grant 2018-T2/TIC-11732A,
Spanish projects RTI2018-102043-B-I00 SCUM and PGC2018-102210-B-I00 BOSCO,
Madrid regional project S2018/TCS-4339 BLOQUES, and
a Ram\'on y Cajal fellowship RYC-2016-20281.

%

%

%
%
%

%

%

%

%
 % end input ./conclusions.tex
 
\balance
\bibliographystyle{ACM-Reference-Format}
\bibliography{biblio}

\clearpage

\appendix
\onecolumn

\section{MemBlockLang Syntax and semantics}\label{appendix:mbl:syntax-and-semantics}
Here, we present the syntax and semantics of the MBL language.
In the following, we assume given an ordered set of blocks $\Blocks=\{a_1,a_1,\dots, a_m\}$ and a value $n \in \Nat$ representing the cache's associativity such that $n < m$.

\para{Syntax}
The syntax of MBL is given in Figure~\ref{figure:mbl:syntax}.

\begin{figure}[h]
\begin{tabular}{llcl}
\multicolumn{4}{l}{\bf Basic Types} \\
\textit{(Blocks)} 	&  $b$		& $\in$ & $\{a_1,a_1,\dots, a_m\}$ \\
\textit{(Tags)} 		&  $t$ 		& $\in$ & $\{?,!\}$  \\
\textit{(Numbers)}	& $k$ & $\in$ & $\Nat$\\\\
\multicolumn{4}{l}{\bf Syntax} \\
\textit{(Queries)} 	&  $q$		& $:=$ & $\emptysequence \mid b \mid (b)t \mid q_1 \cdot q_2 $ \\
\textit{(Expressions)} 	&  $s$ 		& $:=$ & $(q) \mid \{q_1, \ldots, q_l\} \mid @ \mid \_ \mid (s)t \mid $ \\
& & & $s_1 \circ s_2 \mid  (s_1)[s_2]  \mid (s) k$ \\
\end{tabular}
\caption{MBL syntax}\label{figure:mbl:syntax}
\end{figure}

\para{Semantics}
The semantics of an MBL expression $s$ consists of a set of queries $\semantics{s}$ and it is defined as follows:
\begin{itemize}
	\item $\semantics{q} = \{q\}$ where $q$ is a query.
	\item $\semantics{\{q_1, \ldots, q_l\}} = \{q_1, \ldots, q_l\}$ where $\{q_1, \ldots, q_l\}$ is a set of queries.
	\item $\semantics{\texttt{@}} = \{ a_1 \concat a_2 \concat \dots \concat a_n \}$.
	\item $\semantics{\texttt{\_}} = \{a_1, a_2, \dots, a_n\}$.
	\item $\semantics{s \texttt{?}} = \{ a_1? \concat a_2? \concat \ldots \concat a_k? \mid a_1 \concat a_2 \concat \ldots \concat a_k \in \semantics{s} \}$ where $s$ is an MBL expression such that all queries in $\semantics{s}$ do not have tags.
	\item $\semantics{s \texttt{!}} = \{ a_1! \concat a_2! \concat \ldots \concat a_k! \mid a_1 \concat a_2 \concat \ldots \concat a_k \in \semantics{s} \}$ where $s$ is an MBL expression such that all queries in $\semantics{s}$ do not have tags.
	\item $\semantics{s_1 \circ s_2}  = \{s \concat s' \mid s \in \semantics{s_1} \wedge s' \in \semantics{s_2}\}$ where $s_1,s_2$ are MBL expressions.
	\item $\semantics{s_1[s_2]} = \{ s \concat a \mid s \in \semantics{s_1} \wedge \exists s_2 \in \semantics{s_2}, i \in \Nat.\ \element{s_2}{i} = a\}$ where $s_1,s_2$ are MBL expressions.
	\item $\semantics{(s)k} = \{ s_1 \concat s_2 \concat \ldots \concat s_k \mid \bigwedge_{1 \leq i \leq k} s_i \in \semantics{s} \}$, where $s$ is an MBL expression and $k \in \Nat$.
	\item $\semantics{(s)} = \semantics{s}$, where $s$ is an MBL expression.
\end{itemize}

\section{Adaptive Policies and Leader Sets}\label{appendix:leaders}

Henry Wong~\cite{ReplacementPolicy2013} identifies an adaptive L3 cache on Intel's Ivy Bridge processors and describes heuristics for detecting the existence of leader sets.

In a similar way, we use \cachequery{} to run several thrashing queries (i.e., access patterns with a working set that does not fit into the cache and degenerates performance) on a per set basis, obtaining the following results:

\begin{itemize}
	\item Haswell i7-4790: sets $512-575$ in slice $0$ implement a fixed policy susceptible to thrashing. Sets $768-831$ in slice $0$ implement a fixed thrash resistant policy (that seems to be not deterministic). In contrast, the rest of the cache sets follow the policy producing less misses.
	\item Skylake i5-6500: cache sets whose indexes satisfy $((((\mathtt{set} \mathrel{\&} \mathtt{0x3e0}) \gg 5) \oplus (\mathtt{set} \mathrel{\&} \mathtt{0x1f})) = \mathtt{0x0}) \land ((\mathtt{set} \mathrel{\&} \mathtt{0x2}) = \mathtt{0x0})$
	implement a fixed policy susceptible to thrashing (i.e., policy {\em New2}).
	The rest seem to use an adaptive policy that behaves in a non-deterministic way.
	\item Kaby Lake i7-8550U: we observe the same behavior and set selection than on Skylake i5-6500.
\end{itemize}

On Haswell, we confirm previous results reported in \cite{ReloadRefresh2019}. However, we remark that leader sets are only present in slice $0$. It is also worth mentioning that the ranges seem to be selected by comparing the index bits with some fixed constants:
$((\mathtt{set} \mathrel{\&} \mathtt{0x7c0}) \gg 6) = \mathtt{0x8}$
and
$((\mathtt{set} \mathrel{\&} \mathtt{0x7c0}) \gg 6) = \mathtt{0xc}$,
respectively.

Previous work~\cite{eviction19} identified sets with a fixed policy on Skylake, and argued that leader set influence did {\em not} cross slices. We complete this knowledge, including the description for Kaby Lake, and expose that adaptivity {\em has} effects across different slices, i.e. a single cache set leader producing lots of hits can affect all the follower sets in the cache.

We also report the following observations regarding the adaptive policy implemented in the rest of the L3 cache sets in Skylake and Kaby Lake:
\begin{itemize}
\item First, we observe another group of sets ---those whose indexes satisfy
$((((\mathtt{set} \mathrel{\&} \mathtt{0x3e0}) \gg 5) \oplus (\mathtt{set} \mathrel{\&} \mathtt{0x1f})) = \mathtt{0x1f}) \land ((\mathtt{set} \mathrel{\&} \mathtt{0x2}) = \mathtt{0x2})$---
that change at a different rate than the majority.

\item Second, it is possible to control the adaptive policies in 2 ways, by only interacting with the thrash-vulnerable fixed sets: (1) heavily thrashing the fixed sets makes the adaptive policy become more thrash resistant, i.e., \verb!@ M a M?! always produces a miss; (2) continuously producing hits on the fixed sets makes them tend towards the {\em New2} policy.
\end{itemize}
We remark that this interaction needs to happen concurrently, which might indicate a small counter refereeing the set dueling mechanism.
If the unknown adaptive policy is inspired by DRRIP~\cite{Jaleel2010}, it could behave deterministically when completely saturated. However, we have not yet been able to learn it.

Interestingly, the set selection uncovered for Skylake and Kaby Lake processors is very similar to that in~\cite{Qureshi2007}, which augments our confidence in the abovementioned explanation.

\clearpage
\section{Previously Undocumented Policies}\label{appendix:pseudocode}

Figure~\ref{figure:sketchsolutions} shows a cleaned-up version, with minor variable renaming and layout adjustments, of the synthesized explanations describing the 2 previously undocumented cache replacement policies from \S~\ref{sec:evaluation:hardware}.
%
%We include the complete code for all templates, constrains, and solutions, in our repository:
%
%\url{https://github.com/cgvwzq/polca/}.

\begin{figure}[h]
\centering
\begin{minipage}[t]{0.49\textwidth}
\begin{subfigure}[t]{\linewidth}
\begin{lstlisting}[style=CStyle, mathescape]
int[4] s0 = {3,3,3,0};

int[4]hitState (int[4] state, int pos)
  int[4] final = state
  // Promotion
  final[pos] = 0
  // Normalization
  // is there a block with age 3?
  bit found = 0
  for(int j = 0; j < 4; j = j + 1)
    if(!found)
      for(int i = 0; i < 4; i = i + 1)
        if(!found && final[i] == 3)
            found = 1
	// If not, increase all ages
	// except promoted one
    if(!found)
      for(int i = 0; i < 4; i = i + 1)
        if(i != pos)
          final[i] = final[i] + 1
  return final

int missIdx (int[4] state)
  // Replace first block with age 3
  // starting from the left
  for(int i = 0; i < 4; i = i + 1)
    if(state[i] == 3)
      return i

int[4] missState (int[4] state)
  int[4] final = state
  // Insertion
  int replace = missIdx(state);
  final[replace] = 1
  // Normalization
  // Is there a block with age 3?
  bit found = 0
  for(int j = 0; j < 4; j = j + 1)
    if(!found)
      for(int i = 0; i < 4; i = i + 1)
        if(!found && final[i] == 3)
          found = 1
	// If not, increase all ages
	// except inserted one
    if(!found)
      for(int i = 0; i < 4; i = i + 1)
        if(replace != i)
          final[i] = final[i] + 1
  return final
\end{lstlisting}
\caption{Sketch solution for {\em New1} undocumented policy.}
\end{subfigure}
\end{minipage}
\hspace{2pt}
\begin{subfigure}[t]{0.49\textwidth}
\begin{lstlisting}[style=CStyle, mathescape]
int[4] s0 = {3,3,3,3};

int[4] hitState (int[4] state)
  int[4] final = state
  // Promotion
  if(final[pos] < 2 && final[pos] == 1)
    final[pos] = 0
  else if (state[pos] > 1)
    final[pos] = 1
  // Normalization
  // Is there a block with age 3?
  bit found = 0
  for(int j = 0; j < 4; j = j + 1)
    if(found == 0)
      for(int i = 0; i < 4; i = i + 1)
        if(!found0 && final[i] == 3)
          found = 1
    // If not, increase all ages
    if(!found)
      for(int i = 0; i < 4; i = i + 1)
        final[i] = final[i] + 1
  return final

int missIdx (int[4] state)
  // Replace first block with age 3
  // starting from the left
  for(int i = 0; i < 4; i = i + 1)
    if(state[i] == 3)
      return i


int[4]missState (int[4] state)
  int[4] final = state
  // Insertion
  int replace = missIdx(state);
  final[replace] = 1;
  // Normalization
  // Is there a block with age 3?
  bit found = 0
  for(int j = 0; j < 4; j = j + 1)
    if(!found)
      for(int i = 0; i < 4; i = i + 1)
        if(!found && final[i] == 3)
          found = 1
    // If not, increase all ages
    if(!found)
      for(int i = 0; i < 4; i = i + 1)
        final[i] = final[i] + 1
  return final
\end{lstlisting}
\caption{Sketch solution for {\em New2} undocumented policy.}
\end{subfigure}
\caption{Synthesized high-level programs for previously undocumented replacement policies using the {\em Extended} template.}
\label{figure:sketchsolutions}
\end{figure}

\end{document}